# Scalable covalently functionalized black phosphorus hybrids for broad-spectrum virucidal activity

Na Xing[a#], Jasmin Er[ab#], Ricardo M. Vidal[c,d], Sandhya Khadka[a], Robert Schusterbauer[ab], Maik Rosentreter[a], Ranen Etouki[a], Rameez Ahmed[a], Taylor Page[a], Philip Nickl[a], Obida Bawadkji[a], Anja Wiesner[a], Jörg Radnik[b], Vasile-Dan Hodoroaba[b], Kai Ludwig[e], Jakob Trimpert[de], Ievgen S. Donskyi[ab*]

[#] These authors contributed equally

[a]Institut für Chemie und Biochemie, Freie Universität Berlin, Takustraße 3, 14195 Berlin, Germany

[b]Federal Institute for Material Research and Testing (BAM), Division 6.1, Unter den Eichen 44-46, 12205 Berlin, Germany

[c]Institut für Virologie, Freie Universität Berlin, Robert-von-Ostertag-Straße 7-13, 14163 Berlin, Germany,

[d]College of Veterinary Medicine, Kansas State University, Mosier Hall, 1800 Denison Avenue, Manhattan, KS 66506-5600, USA.

[e]Forschungszentrum für Elektronenmikroskopie und Gerätezentrum BioSupraMol, Freie Universität Berlin, Fabeckstraße 36A, 14195 Berlin, Germany

*Corresponding author. E-mail: ievgen.donskyi@fu-berlin.de

At the onset of viral outbreaks, broad-spectrum antiviral materials are crucial before specific therapeutics become available. We report scalable, biodegradable black phosphorus (BP) hybrids that provide mutation-resilient virucidal protection. BP sheets, produced via an optimized mechanochemical process, are covalently functionalized with 2-azido-4,6-dichloro-1,3,5-triazine to form P=N bonds. Fucoidan, a sulfated polysaccharide with intrinsic antiviral activity, and hydrophobic chains are then incorporated to achieve irreversible viral deactivation. The material exhibits strong antiviral inhibition and complete virucidal activity against multiple viruses, including recent severe acute respiratory syndrome coronavirus-2 (SARS-CoV-2) variants. It maintains high biocompatibility, remains effective against viral mutations, and is shelf stable for at least five month. The combination of biodegradability, scalable synthesis, and synergistic antiviral and virucidal mechanisms establishes BP-conjugates as a new class of highly efficient antivirals. They offer a broad spectrum antiviral solutions that could bridge the gap between antiviral medicines and general antiseptics.

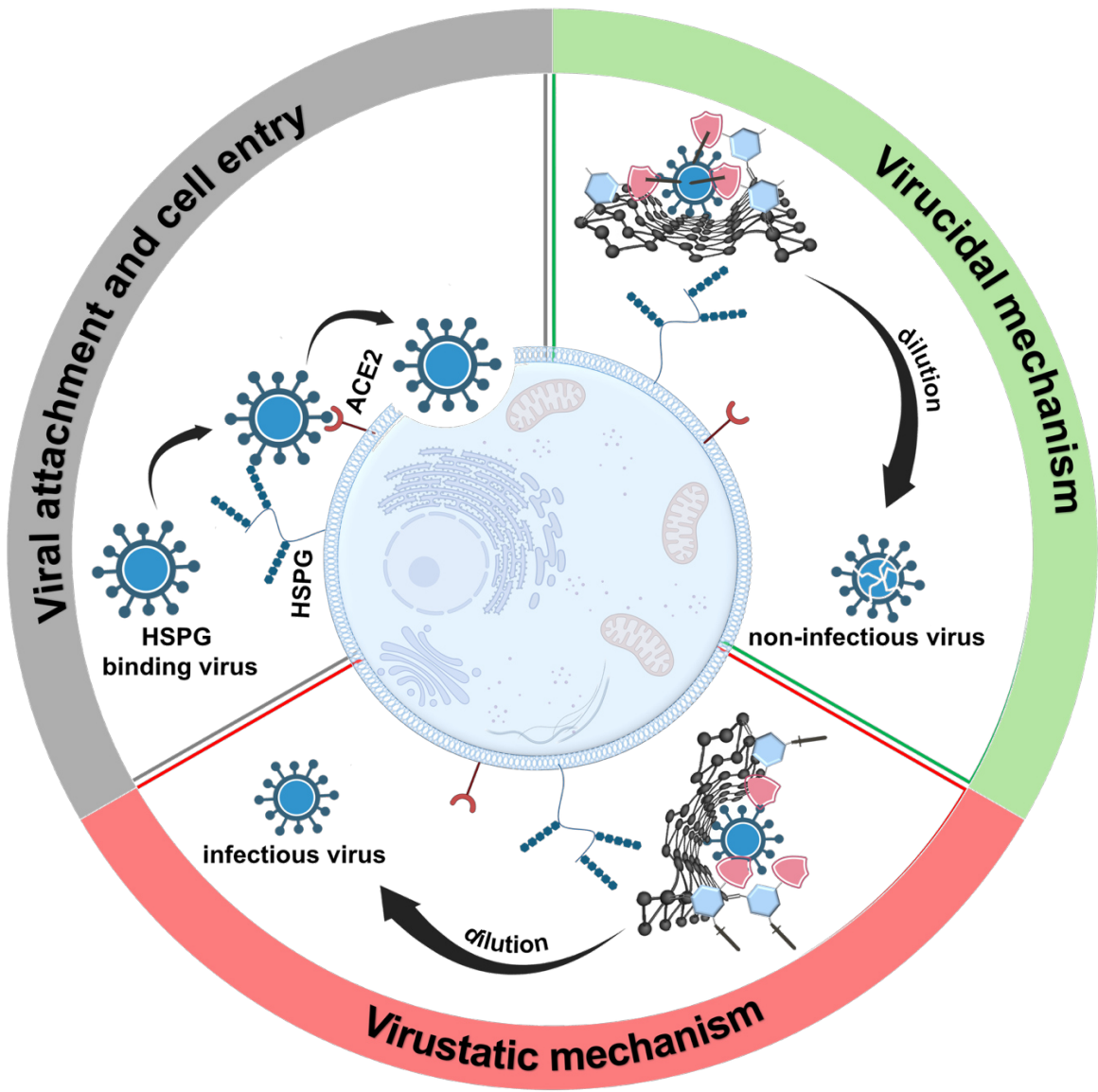

## Introduction

Viral infections remain a significant threat to public health as various epidemics and pandemics have shown throughout the history.[1] The influenza A virus (IAV) pandemic between 1918-1919, claiming more than 40 million lives[2] and the recent pandemic caused by SARS-CoV-2, with over 7 million deaths,[3] underscore the disruptive impact of novel viruses on a population lacking immunity. Current antiviral drugs and vaccines are limited in their use due to their virus-specific nature.[4] In addition, the rapid mutation and emergence of new viral variants, along with resistance to existing antiviral treatments highlight the urgent need for broad-spectrum antiviral agents. These agents should defend against both existing and newly emerging viruses effectively.[4,5] Many viruses use electrostatic interactions for the initial attachment to the host cells. Positively charged amino acids residues on the virus surface bind to negatively charged glycans, such as heparan sulfate proteoglycan (HSPG). HSPG is highly negative proteoglycan that can be found in almost all eukaryotic cells.[6] HSPG serves as a receptor for different viruses, including the human immunodeficiency virus type 1, herpes simplex virus (HSV), respiratory syncytial virus, and SARS-CoV-2.[7–9] HSPG-mimicking materials, including polysaccharides, linear and dendritic polymers and coated nanoparticles, have been developed as potential inhibitors across multiple viral families.[10–12]

Viral inhibition can occur through two distinct mechanisms: virustatic inhibition, where inhibitor binds to the virus preventing cell attachment and therefore shields and hinders its cell entry. Consequently, binding based on electrostatic interactions is reversible and the inhibitory effect is lost upon dilution.[9] Second, virucidal mechanism irreversibly inactivates the virus, rendering it noninfectious. Previous studies demonstrate that the combination of multivalent electrostatic and hydrophobic interactions can result in an virucidal activity against various viruses.[9,13,14]

Fucoidan, a polysaccharide biosynthesized by brown algae, has been recently studied for its diverse biological activities, including anticoagulant, antioxidant, anti-inflammatory and antiviral properties.[2,15,16] Fucoidan extracted from different species differs in its composition, molecular weight, and degree of sulfation.[16] The primary monosaccharide is L-fucose, with varying amounts of other sugars and functional groups.[17] Fucoidan extracted from *Undaria Pinnatifida* (FU) has demonstrated antiviral activity against IAV, HSV type 1 (HSV-1) and 2, human cytomegalovirus and SARS-CoV-2.[17–19]

BP is an emerging two-dimensional (2D) nanomaterial with distinct properties and promising applications in energy storage,[20] catalysis,[21] optoelectronics[22] and biomedicine.[23,24] Unlike the majority of 2D nanomaterials, BP is considered to be biodegradable and biocompatible.[25,26] The phosphorus atoms have a lone ion pairs that enhance the oxophilicity of the material, making it prone to degradation under ambient conditions.[27,28] The degradation of BP nanosheets (BPNS) leads to the conversion of the material into different phosphate anions, including $PO_2^{3}$, $PO_3^{3-}$, $PO_4^{3-}$ species.[29,30] Phosphorus and phosphate anions are highly abundant in living organisms, as part of the DNA backbone, proteins, lipids, and ATP. In the human body, BP is expected to degrade into non-toxic phosphate ions, that are readily metabolized.[31] Compared to other 2D nanomaterials, BP has a higher specific surface area, due to its puckered honeycomb structure, resulting in a higher drug loading capacity and functionalization degree.[32] At the same time, the functionalization of the BP surface with azides, diazonium salts and halides enhance the overall stability of the material, which is crucial for the practical use of the material.[33]

In this work, we present a highly efficient broad-spectrum antiviral material based on degradable black phosphorus-polysaccharide conjugation. FU introduces virus-binding capability, while hydrophobic chains enable irreversible inactivation of multiple viruses. Produced material stays efficient months after its synthesis and retains its activity after 13 cycles of viral mutations. The BP derivatives are comprehensively characterized by X-ray photoelectron spectroscopy (XPS), near-edge X-ray absorption fine structure (NEXAFS), atomic force microscopy (AFM), Raman spectroscopy, cryogenic transmission electron microscopy (cryo-TEM), time-of-flight secondary ion mass spectrometry (ToF-SIMS), and X-ray diffraction (XRD). The study further evaluates the biological responses to these hybrids and demonstrates their ability to both bind and deactivate viruses. The most efficacious material, functionalized with palmitic acid, completely inactivates multiple viruses, including recent SARS-CoV-2 variants, and preserves a broad therapeutic window.

**Results and discussion**

Beginning with red phosphorus (RP), the production of BPNS was optimized via a planetary ball-milling approach, providing an environmentally friendly and scalable method without requiring toxic catalysts (i.e. tin, iodine).[34] The conversion of RP to BP (Fig. 1A) was monitored using Raman spectroscopy (Fig. 1B) and XRD (Fig. 1C). Compared to RP, BP had three characteristic peaks in Raman spectroscopy at ~360 cm$^{-1}$, ~430 cm$^{-1}$ and ~460 cm$^{-1}$ corresponding to the $A_g^1$, $B_{2g}$ and $A_g^2$ vibrational modes respectively. Raman spectroscopy, XRD and UV/Vis spectroscopy collectively indicated nearly complete conversion after 30 min of milling with the reverse mode activated (30 min R). The high-resolution P2p XPS spectrum of pristine RP indicated partial oxidation indicated by the P-O peak at 134.9 eV eV (Fig. 1D). Interestingly, the converted BP revealed no sign of oxidation, suggesting that the P-O bonds were cleaved during the ball-milling process (Fig. 1E). Next, BPNS were prepared via liquid-phase exfoliation in *N*-methyl-2-pyrrolidone (NMP) using sonication. Bath sonication for 8 h yielded in nanosheets with an average thickness of ~100 nm and lateral dimensions of 300-800 nm (Fig. 1F). However, the production process was limited by an average yield of 5-10 mg. In stark contrast, probe sonication reduced the optimal exfoliation time to 1h, while significantly enhancing both quality and material output. The resulting BPNS exhibited a decreased average thickness of ~50 nm with lateral sizes of 1-2 μm and a markedly improved yield between 100-150 mg (Fig. 1G). The high resolution P2p spectra for both methods showed no sign of oxidation (Fig. 1H and 1I) and the absence of nitrogen signals in the survey spectrum confirmed efficient removal of NMP (Fig. S1). The stability of bulk BP and the exfoliated BPNS were systematically investigated by XPS, (Fig. S2) where samples were stored under ambient conditions and high-resolution P2p spectra were recorded daily. Bulk BP displayed rapid oxidation after 2 d, increasing to ~50% after 4 d and nearly complete surface oxidation after 7 d (Fig. S2). In contrast, both BPNS (exfoliated via bath and probe sonication) revealed remarkable stability, with minimal oxidation detected after 7d of exposure to ambient conditions. Next, the feasibility of an environmentally sustainable aqueous-based approach was investigated. When degassed water was employed as the exfoliation solvent, probe sonication (1h) yielded ultrathin nanosheets with an average thickness of ~3 nm and lateral dimension of 100-300 nm (Fig. S3I). However, the high resolution P2p spectrum revealed slight oxidation at 135.2 eV (Fig. S3C). Therefore, further functionalization steps were discussed for non-oxidized NMP exfoliated BPNS.

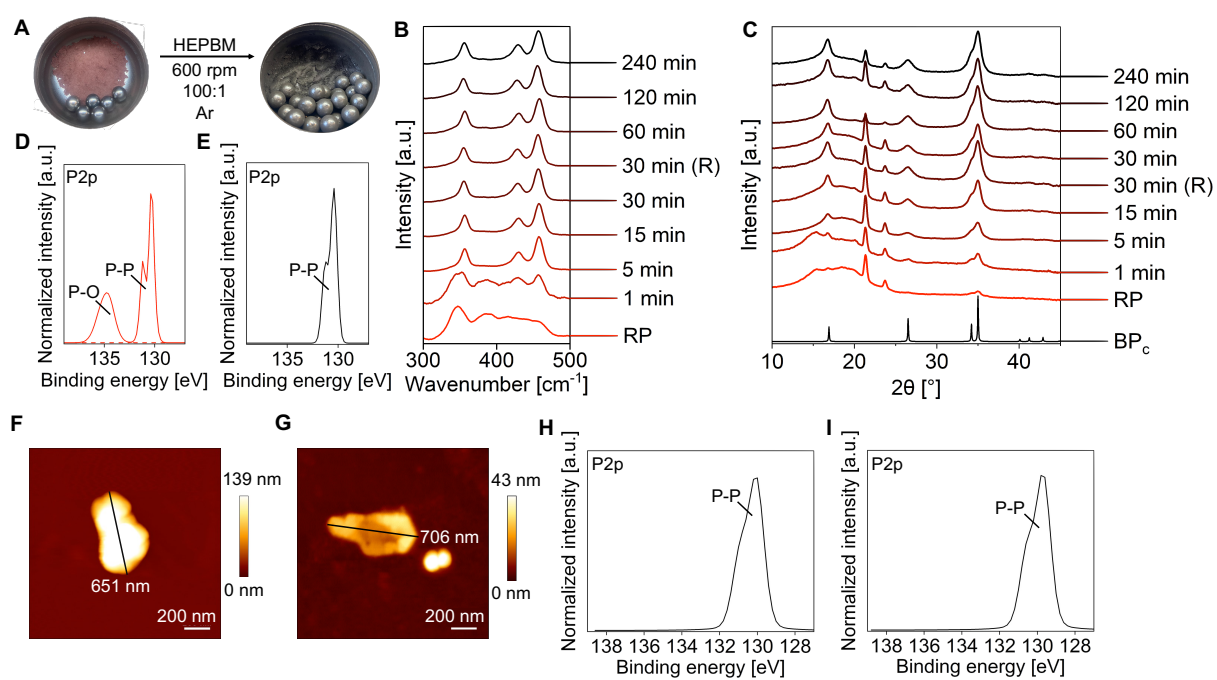

**Figure 1.** (A) Optimized conditions for the conversion of RP to BP, with images of the ball-mill chamber showing optical transition from red to black. (B) Raman spectra for the conversion of RP to BP at different milling times, showing appearance of the characteristic BP peaks ($A_g^1$, $B_{2g}$ and $A_g^2$) after 5 min of milling. (C) XRD patterns for the conversion of RP to BP at different milling times, showing full formation of the characteristic BP peaks ((020), (021), (040) and (111)) after 30 min of milling with the reverse mode on (30 min R) and comparison to crystalline BP ($BP_c$). High resolution P2p XPS spectra for (D) RP, showing slight oxidation and (E) BP showing no sign of oxidation. AFM images of (F) BPNS exfoliated in NMP via bath sonication (the scale bar corresponds to 200 nm) and (G) BPNS exfoliated in NMP via probe sonication (the scale bar corresponds to 200 nm). High resolution P2p XPS spectra for (H) BPNS exfoliated in NMP via bath sonication and (I) BPNS exfoliated in NMP via probe sonication.

BPNS were covalently functionalized via a nitrene [2+1] cycloaddition reaction, enabling the conjugation of the 4,6-dichloro-1,3,5-triazine moiety to the surface (Fig. 2A). This strategy represented a significant advancement in the chemical modification and stability of BPNS, introducing robust covalent bonds and the possibility for dual post-modification. Comprehensive spectroscopic analyses confirmed successful functionalization. Fourier transform infrared (FTIR) spectroscopy of 2-azido-4,6-dichloro-1,3,5-triazine exhibited a characteristic -$N_3$ peak at 2169 cm$^{-1}$. BPNS showed weak signals at 1201 cm$^{-1}$ (P=O), 993 cm$^{-1}$ and 900 cm$^{-1}$ (P-O), arising from minor surface oxidation during work-up (Fig. 2K). Upon functionalization with 2-azido-4,6-dichloro-1,3,5-triazine, the formation of a new peak at 1013 cm$^{-1}$ was observed, indicating the formation of P-N bonds[35] and showing successful functionalization. While the functionalized BPNS-Trz retained three characteristic vibrational modes of BP in Raman spectroscopy ($A^1_g$, $B_{2g}$ and $A^2_g$), a slight red-shift to lower energies was observed (Fig. 2H), due to the hinderance in oscillation of the covalently functionalized phosphorus atoms, in agreement with previous reports on covalently functionalized BP.[36] XPS data provided direct evidence of the successful incorporation of the triazine moiety and allowed for a quantitative assessment of the functionalization degree. A clear nitrogen signal of ~3 % emerged in the survey spectrum of BPNS-Trz, that was absent in unfunctionalized BPNS (Fig. S1), indicating one triazine functionalization per 100 P atoms. To increase the functionalization degree, tetrabutylammonium bromide was used as a phase-transfer catalyst, resulting in an increase in N content to 7 %, corresponding to one triazine group per ~50 phosphorus atoms (Table S3). In the high resolution C1s spectra of both materials the three peaks at 285.5 eV, 287.0 eV and 289.2 eV corresponded to the C-N=C, C-Cl and C-N-P bonds of the dichlorotriazine groups, respectively (Fig. 2D and 2G). The high-resolution N1s XPS spectra exhibited two peaks, one at 399.3 eV attributed to the P-N bonds formed on BPNS surface and another one at 401.9 eV corresponding to the N-C=N bonds within the triazine ring (Fig. 2C and 2F). The relative peak areas of 25% and 75% aligned with the stoichiometric distribution of three nitrogen atoms in the triazine ring and the single nitrogen atom covalently bonded to the BPNS. Remarkably, the high resolution C1s and N1s spectra for the BPNS-Trz (with and without the catalyst) confirmed the same functionalization strategy and gave highly repetitive results but they did not give information about the functionalization degree. The high resolution P2p spectra offered a more direct measure of functionalization degree. For both materials three peaks were observed. The P-P bond at 130.1 eV, the P-N bond at 133.4 eV and some oxidized P-$O_x$ species at 135.0 eV respectively (Fig. 2B and 2E). For the sample prepared without the catalyst, the P-N peak was weak and broad (Fig. 2B), suggesting a lower functionalization degree. The P-N peak for the sample prepared with the catalyst exhibited a more intense and defined peak confirming the improved functionalization degree (Fig. 2E). This enhancement underlined the efficacy of tetrabutylammonium bromide as a phase-transfer catalyst significantly increasing the functionalization degree of BPNS with dichlorotriazine. Further, hard X-ray photoelectron spectroscopy (HAXPES) was performed on the functionalized material. Unlike conventional XPS (information depth ~10 nm), HAXPES allowed access to information from deeper layers and the detection of more electron levels like the P1s. For the high resolution P1s spectrum a clear signal was observed for the P-P bonds at 2142.5 eV and a slight signal at 2147.0 eV corresponding to minor oxidation (Fig. S4A). However, no clear signals were detected for the N1s and C1s core levels (Fig. S4B and C). This absence confirmed that the triazine moieties were located on the surface of the BPNS and that covalent functionalization occurred on the outermost layers. To further confirm the successful functionalization and to elucidate the binding environment NEXAFS studies were performed at the carbon and nitrogen K-edges. The NEXAFS C k-edge spectrum of BPNS-Trz revealed several resonance features related to the dichlorotriazine moiety. A pronounced resonance at 289.5 eV was attributed to the C1s $\pi^*$ transition of the sp$^2$-hybridized C=N bonds. Another sharp resonance feature was observed at slightly higher binding energies (290.3 eV) referred to the C1s $\sigma^*$ transition of C-N bonds. In the higher energy region between 292.6 eV and 296.7 eV various overlapping resonance features aroused as the C1s $\sigma^*$ transitions of the dichlorotriazine moiety. The most significant feature at 295.5 eV was associated with the C1s $\sigma^*$ (C-Cl) transition (Fig. 2I). At the N K-edge, the NEXAFS spectrum exhibited two main peaks at 399.0 eV and 400.8 eV assigned to N1s $\pi^*$ (N=C) and (N=C-Cl) resonances respectively, corresponding to the aromatic nitrogen species within the dichlorotriazine moiety. In addition, another resonance feature at 405.8 eV attributed to the N1s $\sigma^*$ (N-P) transition, confirms the covalent functionalization of the BPNS (Fig. 2J). These results were consistent with previous studies on dichlorotriazine-functionalized graphene, further validating the covalent functionalization strategy.[37] AFM revealed a uniform surface morphology with heights between 10-80 nm and lateral dimension of 300-1000 nm (Fig. 2N). Next, ToF-SIMS supported the presence and spatial distribution of triazine-derived fragments. Both BPNS and BPNS-Trz exhibited a fragment corresponding to oxidized $PO_2^-$ on the surface (Fig. 2L and M). a distinct fragment, identified as $CN^-$, was detected in the functionalized BPNS-Trz exclusively (Fig. 2O and P). In contrast, a distinct fragment at m/z corresponding to $CN^-$, characteristic of the triazine moiety, was detected exclusively for BPNS-Trz (Fig. 2O, P). The $CN^-$ signal was homogeneously distributed across the analyzed area, indicating uniform surface functionalization after the covalent conjugation of triazine moieties. SEM-EDX mapping further support a homogenous surface (covering), revealing consistent P,C and N signals throughout the BPNS-Trz surface (Fig. 2Q).

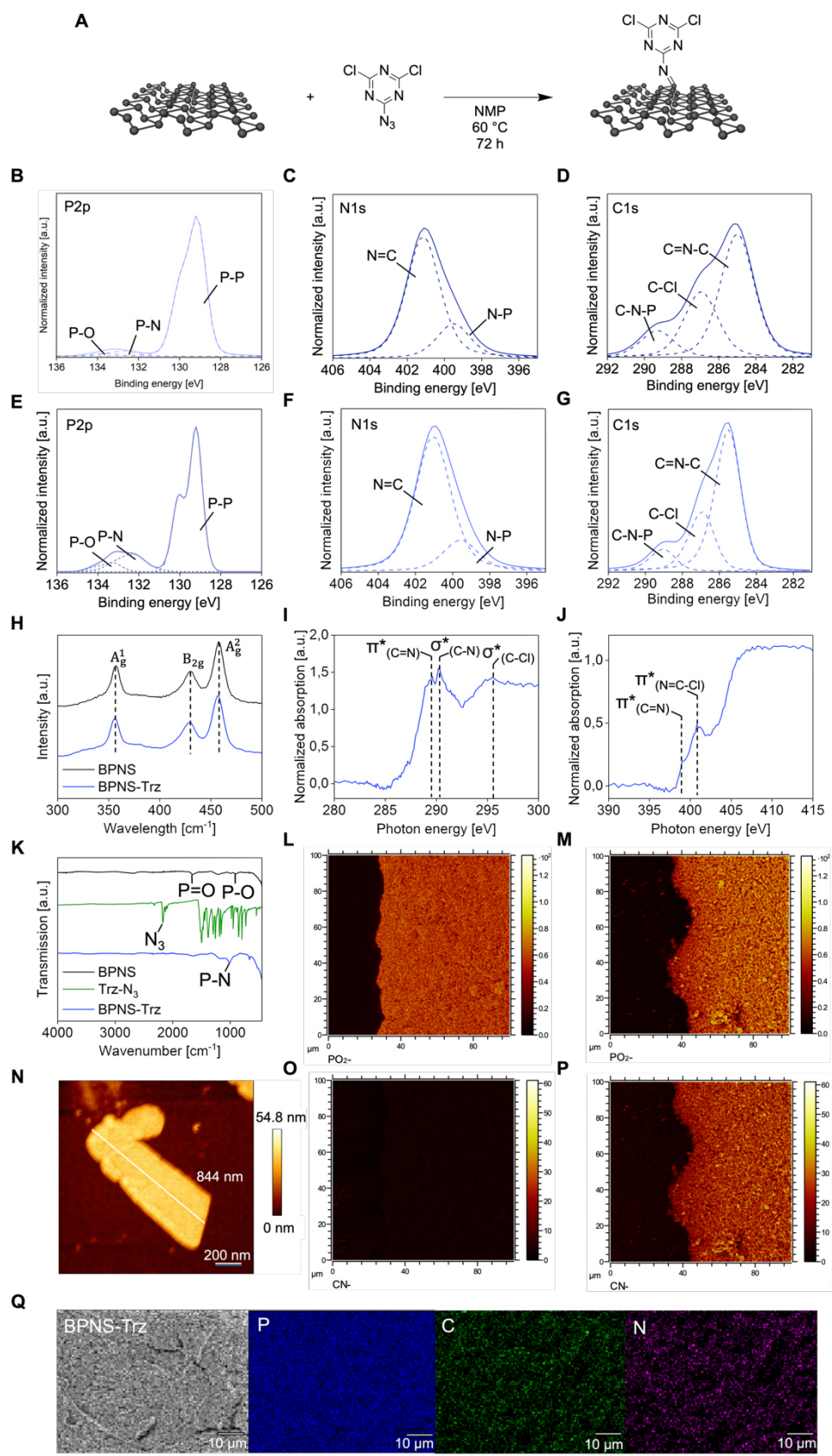

**Figure 2.** (A) Schematic representation of the synthesis of BPNS-Trz using 2-azido-4,6-dichloro-1,3,5-triazine. high resolution XPS spectra with peak fitting for BPNS-Trz without the catalyst (B) P2p, (C) N1s and (D) C1s.

high resolution XPS spectra with peak fitting for BPNS-Trz with TBAB (E) P2p, (F) N1s and (G) C1s. (H) Raman comparison of BPNS (black) and BPNS-Trz (blue). NEXAFS spectrum of BP-Trz at the (I) C K-edge and (J) N K-edge, respectively. (K) IR comparison of BPNS (black), Trz-N$_3$ (green) and BPNS-Trz (blue) with characteristic peaks. (N) exemplary AFM image of BPNS-Trz. ToF-SIMS of PO$^{2-}$ fragment in the negative mode of (L) BPNS and (M) BPNS-Trz. ToF-SIMS of CN$^-$ fragment in the negative mode of (O) BPNS and (P) BPNS-Trz. (Q) SEM-EDX mapping of the surface of BPNS-Trz (the scale bar corresponds to 10 μm).

To understanding the nitrene functionalization of BPNS surfaces, two binding motifs were previously suggested by Liu et al.[38] were considered. One binding mode resulted in a species similar to iminophosphoranes with a P=N double bond between the nitrene and the phosphorus (Fig. 3A (isomer 2)). The other binding mode resulted in a disphosphine amine bridge between the nitrene and two neighboring phosphorus atoms leading to the cleavage of a P-P bond (Fig. 3B (isomer 1)). Reference compounds were synthesized with both binding motifs, following modified protocols for the formation of different iminophosphoranes[39] (Fig. 3C) and diphosphine amines[40] (Fig. 3D). Here, all reference materials and BPNS-Trz exhibited two distinct components, corresponding to carbon-nitrogen bonds and phosphorus-nitrogen bond.

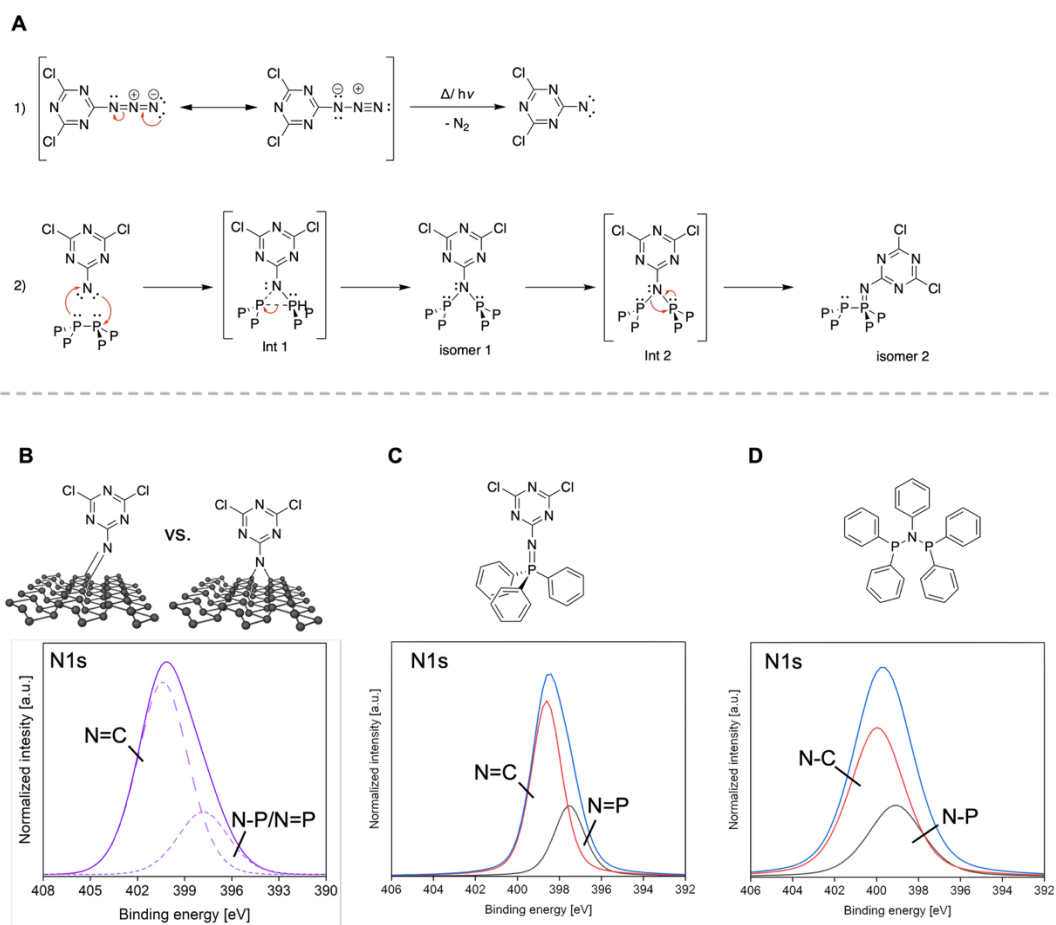

**Figure 3.** (A) Proposed mechanism for the reaction of BPNS with dichlorotriazine.[38] (B) High-resolution N1s spectra of BPNS-Trz with both binding modes. (C) High-resolution N1s of the reference compound with a P=N bond. (D) High-resolution N1s of the reference compound with a P-N bond.

To ensure virus binding, FU was used as a heparan mimic. Aminated FU (FU-NH$_2$) was covalently attached to BPNS-Trz via a nucleophilic aromatic substitution of a chlorine atom under mild conditions at room temperature, yielding a water-soluble product (BPNS-Trz-FU) (Fig. 4A). The high-resolution C1s revealed characteristic peaks at 285.0 eV for the C-C component, 286.2 eV for C-N and C-O components and 287.7 eV for the C=O component, respectively (Fig. 4B). The high resolution N1s spectrum revealed the expected component for amination at 401.0 eV (Fig. 4C). Due to complete surface coverage by FU-NH$_2$ (Fig. 4J), no phosphorus was detected in XPS (Fig. 4D). Further, we used hard XPS (HAXPES) to obtain deeper insight into the bonding environment. In the high resolution P1s spectrum a signal for P-P (2142.5 eV) and two signals for P-N (2144.0 eV) and P-O (2147.6 eV) were observed (Fig. 4G). Owing to the increased information depth of HAXPES underlying BP remained detectable in contrast to XPS. In the high-resolution HAXPES C1s (Fig. 4E) and N1s (Fig. 4F) spectra, no additional signals were observed. These findings confirmed that the FU formed a covalent bond to the BPNS-triazine surface. FU is FTIR active and showed all characteristic peaks, including an OH band at 3485 cm$^{-1}$, C-H vibrations at 2980 cm$^{-1}$, C=O ester stretching at 1636 cm$^{-1}$, S=O stretching and deformation at 1220 cm$^{-1}$ and 580 cm$^{-1}$ respectively, and C-O ether stretching at 1050 cm$^{-1}$ (Fig. 4I). To further analyze the binding environment NEXAFS studies were performed at the carbon K-edge. The NEXAFS C K-edge spectrum of BPNS-Trz-FU demonstrated several distinct resonance features corresponding to the FU moiety (Fig. 4H). A peak at 288.2 eV was attributed to the C1s π* transition of the sp$^2$-hybridized C=O bonds. In the higher energy region between 292.0 eV and 296.0 eV several overlapping resonance features were observed identified as the C1s σ* transitions of the FU moiety. At the same time, Raman spectroscopy revealed three characteristic peaks for BP (Fig. S6A) confirming that BP was still detectable and present in BPNS-Trz-FU.

To ensure the virucidal activity of the material, aliphatic chains were introduced into the structure (Fig. 4A). Initially, the design strategy involved direct attachment of the aliphatic chains to the triazine ring by substitution of the second chlorine atom. Here, primary alkylamines of 11 and 14 alkyl chains were used (Fig. 4A(i)). This approach was inspired by previous finding, showing that aliphatic chains with >9 carbon atoms show better antiviral properties.[41] FTIR, TGA and elemental analysis confirmed the successful conjugation (Fig. S5 and Table S2). However, initial viral testing revealed that these materials demonstrate strong inhibition but exhibited no virucidality. To overcome this problem, the design was modified, by positioning the aliphatic chains on the FU moiety instead (Fig. 4A(ii)). BPNS-Trz-FU was further functionalized with saturated fatty acids of two different lengths, following a modified protocol from Seiser et al.[42] Capric acid ($C_{10}$) and palmitic acid ($C_{16}$) were chosen, since they are the most abundant saturated fatty acids in the human body.[42] The successful functionalization was indicated by FTIR and NEXAFS. FTIR showed a stronger band at ~2900 cm$^{-1}$ corresponding to the C-H stretching vibration from the aliphatic chains (Fig. 4I). Upon functionalization with palmitic acid, the NEXAFS C K-edge spectrum of BPNS-Trz-FU-$C_{16}$ displayed an additional feature at 289.0 eV. This corresponded to the C1s π* transition associated with the amide C=O bonds (Fig. 4H). AFM analysis revealed material heights of 100-300 nm with smooth morphology and lateral dimension of 500-1000 nm (Fig. 4J). ToF-SIMS revealed no detectable $PO_2^-$ signals (Fig. 4K), but a new fragment corresponding to $SO_3^-$, assigned to the sulfate groups in the FU (Fig. 4M). SEM-EDX elemental mapping revealed a uniform distribution of C, O, N, P and S over the BPNS surface, confirming the successful and homogeneous conjugation of FU (Fig. 4N). Zeta-potential measurements showed consistently negative values across all samples (Table 1), with BPNS-Trz-FU-$C_{16}$ exhibiting a potential of -43.8 mV. This data supports the covalent conjugation of BPNS-Trz-FU and acid-terminated long alkyl chains since both components possess intrinsic negative charges that would lead to electrostatic repulsion.

**Table 1**. Characterization of BPNS derivatives with zeta potential values taken at 1.0 mg/mL, $CC_{50}$ values for the cytotoxicity on Vero E6 cells, $IC_{50}$ for the pre-inhibition of 229E, FCoV and HSV-1.

| Compound | ζ [mV] | Cytotox. $CC_{50}$ [μg/mL] | HSV-1: $IC_{50}$ [μg/mL] | 229E: $IC_{50}$ [μg/mL] | FCoV: $IC_{50}$ [μg/mL] |
|---|---|---|---|---|---|
| **BPNS-Trz-FU** | -40.5 | 275.37±50.57 | 0.51±0.08 | 0.34±0.18 | 10.56±2.88 |
| **BPNS-Trz-FU-$C_{10}$** | -42.3 | ≥1000 | 0.22±0.11 | 0.61±0.08 | 20.80±10.30 |
| **BPNS-Trz-FU-$C_{16}$** | -35.2 | 284.07±50.13 | 0.17±0.04 | 0.30±5.30 | 18.28±6.35 |

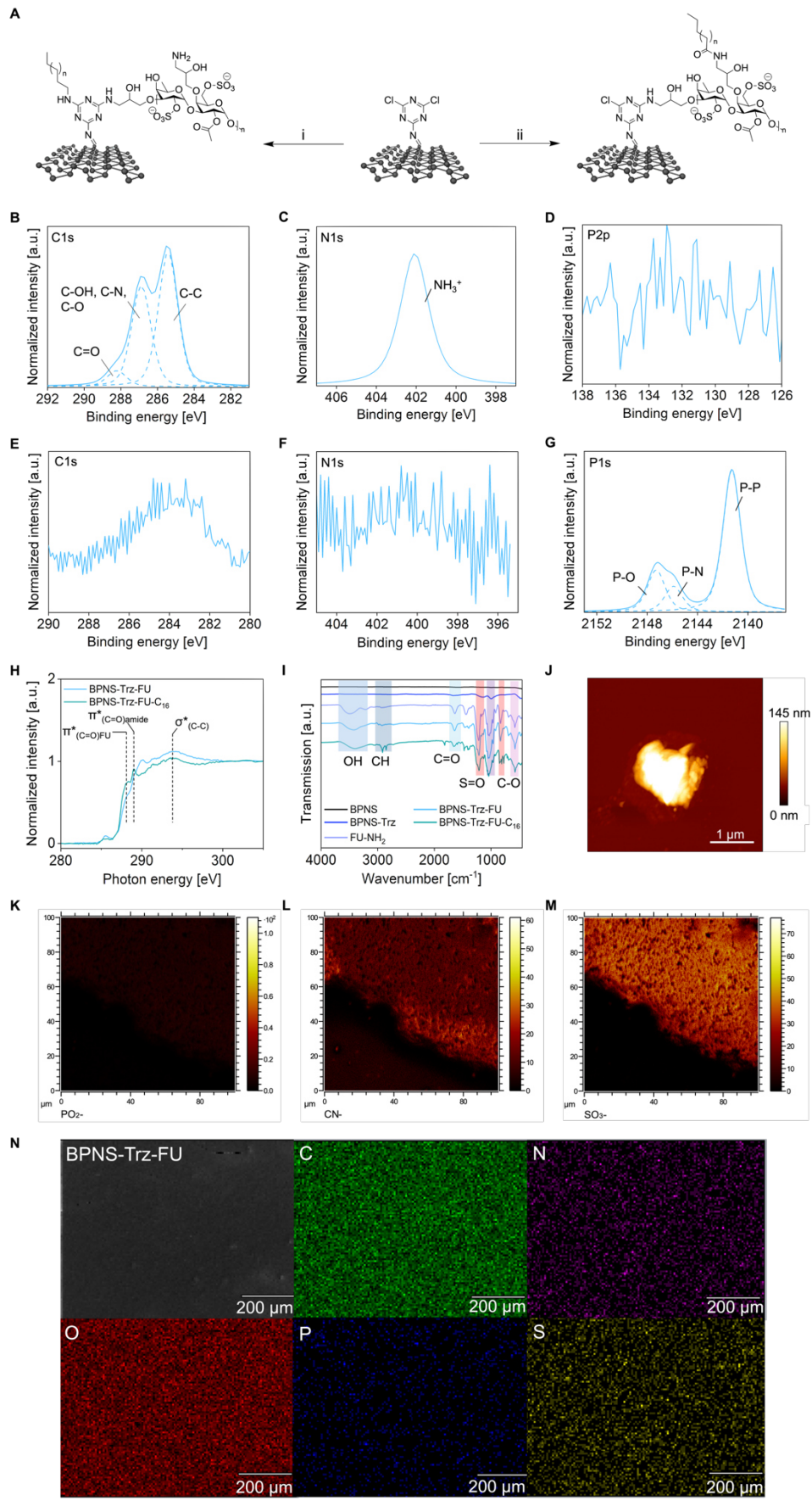

**Figure 4.** (A) Schematic representation of both design strategies (i) BPNS-Trz-FU/$C_x$ with direct attachment of the aliphatic chain to the triazine ring and (ii) BPNS-Trz-FU-$C_x$ with conjugation of the aliphatic chain to the FU

moiety. (B) High resolution XPS spectra with peak fitting for BPNS-Trz-FU (B) C1s, (C) N1s and (D) P2p. (C) High resolution HAXPES spectra with peak fitting for BPNS-Trz-FU (E) C1s, (F) N1s and (G) P2p. (H) NEXAFS comparison at the C K-edge of BPNS-Trz-FU (blue) and BPNS-Trz-FU-$C_{16}$ (green). (I) FTIR comparison of BPNS, BPNS-Trz, FU-$NH_2$, BPNS-Trz-FU, and BPNS-Trz-FU-$C_{16}$. (J) AFM image of BPNS-Trz-FU-$C_{16}$ (the scale bar corresponds to 1 µm). ToF-SIMS in the negative mode of BPNS-Trz-FU-$C_{16}$ fragments (K) $PO_2^-$, (L) $CN^-$ and (M) $SO_3^-$. (N) SEM-EDX element mapping of BPNS-Trz-FU-$C_{16}$ for C, O, N, P, S.

The antiviral efficacy of FU-functionalized BPNS was evaluated against a variety of DNA and RNA viruses, including HSV-1, human coronavirus 229E (HCoV-229E), feline coronavirus (FCoV) and SARS-CoV-2. For HSV-1, all functionalized BPNS compounds displayed potent inhibition, with half-maximal inhibitory concentration ($IC_{50}$) values in the sub-microgram per milliliter range (Table 1). $C_{10}$- and $C_{16}$-chain–modified derivatives, BPNS-Trz-FU-$C_{10}$ and BPNS-Trz-FU-$C_{16}$, were the most active, with $IC_{50}$ values of $0.22 \pm 0.11$ µg/mL and $0.17 \pm 0.04$ µg/mL, respectively, outperforming the non-alkylated BPNS-Trz-FU ($IC_{50}$ = $0.51 \pm 0.08$ µg/mL). Similarly, against HCoV-229E, all compounds showed strong antiviral activity with no statistically significant differences among them. $IC_{50}$ values were $0.34 \pm 0.18$ µg/mL for BP-Trz-FU, $0.61 \pm 0.08$ µg/mL for BPNS-Trz-FU-$C_{10}$, and $0.30 \pm 5.30$ µg/mL for BPNS-Trz-FU-$C_{16}$. (Table 1 and Fig. 5A-C). In the case of FCoV, all derivatives demonstrated significant protection to host cells, however the antiviral activity was reduced compared to HSV-1 and HCoV-229E (Table 1). This diminished efficacy may be attributed to differences in the viral entry mechanisms between coronaviruses. While both HCoV-229E and FCoV belong to the same *Alphacoronavirus* genus, structural variations in their spike proteins, particularly within the receptor-binding domains, could influence their affinity for sulfated decoys. FCoV may exhibit lower dependence on heparan sulfate proteoglycans (HSPGs) for initial attachment and instead rely more on specific receptor interactions (e.g., with feline aminopeptidase N) that are less effectively blocked by broad-spectrum inhibitors.

Further, functionalized BPNS materials were assessed against three Omicron subvariants of SARS-CoV-2: BA.5, EG.5.15, and JN.1. BPNS-Trz-FU selectively inhibited the most recent and epidemiologically relevant subvariants, JN.1, which has rapidly emerged as globally prevalent strain, but revealed no inhibition against either BA.5 and EG.5.15, even at the highest concentrations (Fig. 5E-I).

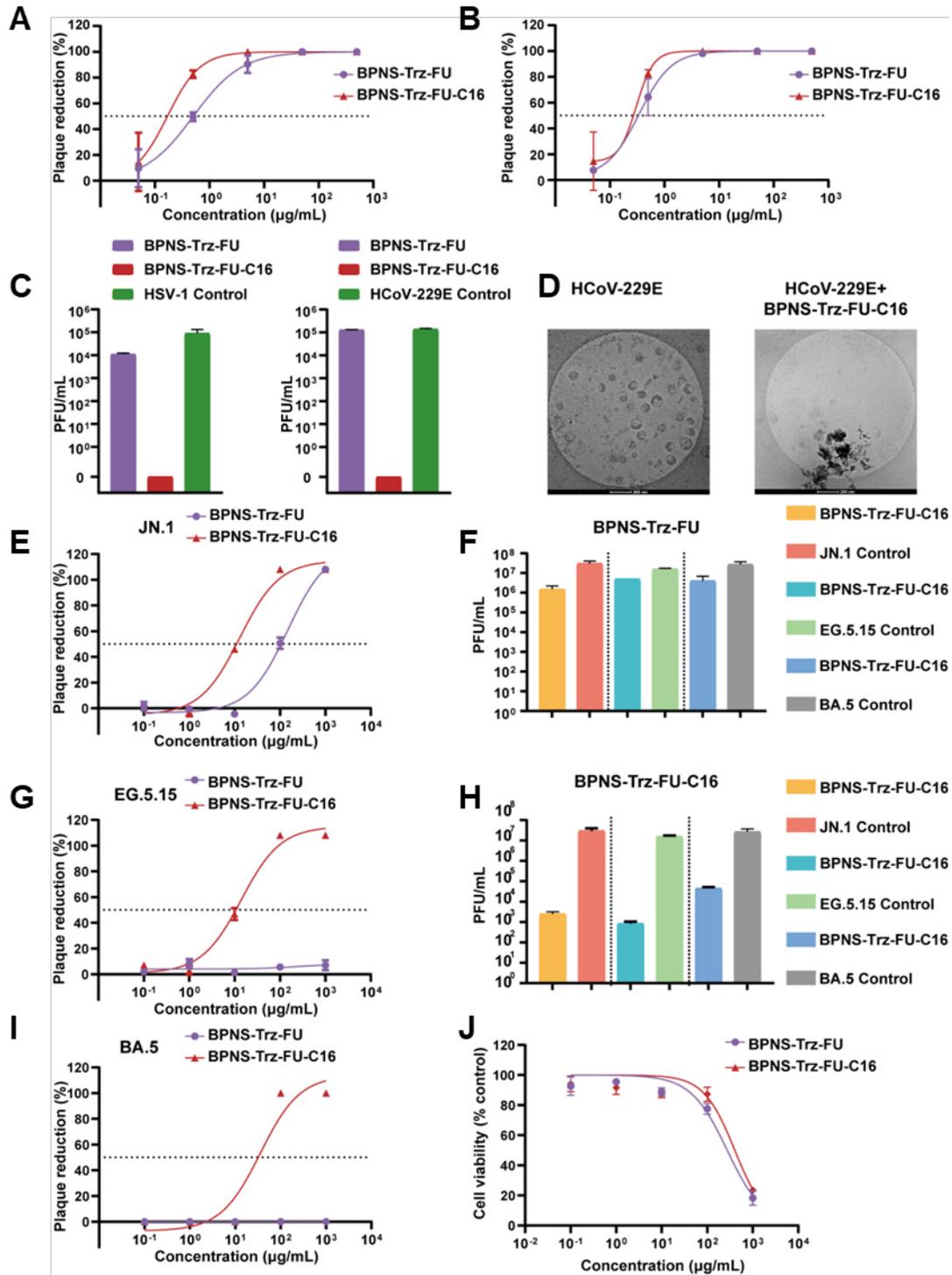

**Figure 5.** (A) HSV-1 and (B) HCoV-229E plaque reduction assay for IC$_{50}$ determination of BPNS-Trz-FU, and BPNS-Trz-FU-C$_{16}$, respectively. (C) HSV-1 and HCoV-229E virucidal assay of BPNS-Trz-FU and BPNS-Trz-FU-C$_{16}$. (D) Cryo-TEM image of purified 229E and a mixture 229E and BPNS-Trz-FU-C$_{16}$ of at concentrations of 5 mg/mL. The scalebar corresponds to 200 nm. (E) JN.1 plaque reduction assay using Vero E6 cells for IC$_{50}$ determination of BPNS-Trz-FU and BPNS-Trz-FU-C$_{16}$. (F) Virucidal assay BPNS-Trz-FU using Vero E6 cells against JN.1, EG5.15 and BA.5. (G) EG.5.15 plaque reduction assay using Vero E6 cells for IC$_{50}$ determination of BPNS-Trz-FU and BPNS-Trz-FU-C$_{16}$. (H) Virucidal assay BPNS-Trz-FU-C$_{16}$ using Vero E6 cells against JN.1, EG5.15 and BA.5. (I) BA.5 plaque reduction assay using Vero E6 cells for IC$_{50}$ determination of BPNS-Trz-FUand

BPNS-Trz-FU-C$_{16}$. (J) Cytotoxicity tested on Vero E6 cells at concentrations ranging from 1000 μg/mL to 0.1 μg/mL. of BPNS-Trz-FU and BPNS-Trz-FU-C$_{16}$.

The observed selectivity of BPNS-Trz-FU for the JN.1 subvariant, while ineffective against BA.5 and EG.5.15, likely stems from continued evolution of the viral spike protein. JN.1 exhibits enhanced dependence on heparan sulfate proteoglycan (HSPG)-mediated attachment due to unique spike mutations (L455S and F456L) that increase positive electrostatic potential, making it highly susceptible to inhibition by sulfated FU-functionalized BPNS. In contrast, earlier subvariants such as BA.5 and EG.5.15 rely more on direct ACE2 binding and less on HSPG interactions. Changes in surface charge and geometry likely limit binding accessibility and reduce the effectiveness of electrostatic inhibitors. Interestingly, BPNS-Trz-FU-C$_{16}$ demonstrated strong inhibitory activity across all three Omicron subvariants (Fig. 5I-K). The compound effectively suppressed viral replication of BA.5, EG.5.15, and JN.1. The higher efficacy of the alkylated derivatives likely results from stronger hydrophobic interactions and multivalent binding with the viral spike protein. Together, these interactions allow BPNS-Trz-FU-C16 to maintain activity against BA.5 and EG.5.15, despite their lower HSPG reliance and modified spike structure. As a result, the material demonstrated broad inhibition efficacy across different viruses and variants.

To investigate the virucidal potential of BP-Trz-FU-C$_{16}$ viruses (HSV-1, HCoV-229E, FCoV, and SARS-CoV-2) were incubated with the compounds for 45 minutes. Following incubation, the samples were diluted to concentrations 10-fold below their IC$_{50}$ values, and residual viral infectivity was quantified using plaque assays. BPNS-Trz-FU-C$_{16}$ demonstrated strong virucidal activity against HSV-1, HCoV-229E (Fig. 5C), and multiple Omicron subvariants (including BA.5, EG.5.15, and JN.1) (Fig. 5H), reducing infectious viral titers over 3 orders of magnitude (Fig. 5E-F, L-M). This pronounced inactivation suggests that the incorporation of aliphatic chains is critical for disrupting viral envelope integrity. Cryo-electron microscopy clearly shows the structures of virions alone as well as significant differences in the contrast between BP-derivatives and viruses (Fig. 5G-H). Taken together, these results suggest that hydrophobic BPNS derivatives bearing aliphatic chains can irreversibly inactivate a broad range of viruses, through a combination of electrostatic and hydrophobic interactions.

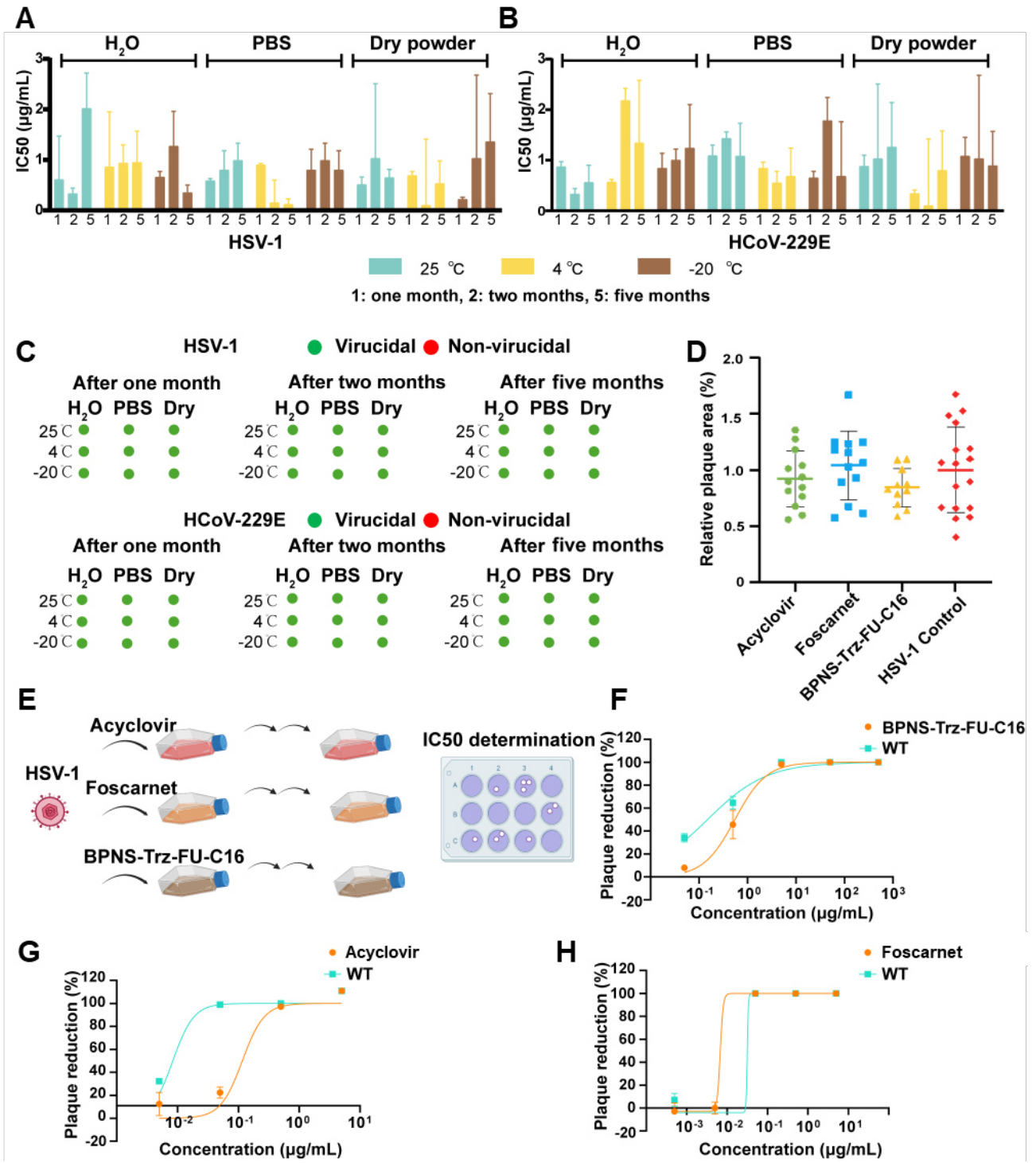

**Figure 6.** IC$_{50}$ values of BPNS-Trz-FU-C$_{16}$ against HSV-1 (A) and HCoV-229E (B) after 1, 2, and 5 months of storage in H$_2$O, PBS, or dry powder at 25 °C, 4 °C, and −20 °C. Error bars represent standard deviation (n = 3). (C) Summary of virucidal (green dots) versus non-virucidal (red dots) behavior after 1, 2, and 5 months of storage under the same conditions as in (A-B). (D) Relative plaque area (%) of HSV-1 after treatment with acyclovir, foscarnet, or BPNS-Trz-FU-C$_{16}$ and untreated control. Error bars represent standard deviation (n = 3). (E) Schematic representation of serial viral passaging (P1 to P13) for HSV-1 in the presence of Acyclovir, Foscarnet, or BPNS-Trz-FU-C$_{16}$, followed by IC50 determination. Mutation-resilient antiviral efficacy as plaque reduction curves of (F) BPNS-Trz-FU-C16, (G) Acyclovir, and (H) Foscarnet against wild-type HSV-1 after 13 serial passaging. Error bars represent standard deviation (n = 3).

The cytotoxicity of FU-functionalized BPNS was assessed using Vero E6 cells. Cells were exposed to concentrations ranging from 0.1 to 1000 µg/mL for 24 hours, and viability was quantified using a luminescent ATP-based assay. All BPNS derivatives exhibited low cytotoxicity at concentrations up to 100 µg/mL, with cell viability remaining above 80% (Fig. 5J). A concentration-dependent decrease in viability was observed at higher doses, culminating in a reduction to below 50% at 1 mg/mL. The effective antiviral concentrations ($IC_{50}$ values in the range of 0.17-0.61 µg/mL for HSV-1 and HCoV-229E) are orders of magnitude lower than the cytotoxic threshold, indicating a wide therapeutic window. These findings suggest that the functionalized BPNS derivatives are biocompatible within their pharmacologically relevant concentration range. The marginal differences in cytotoxicity among BPNS derivatives indicate that alkyl chain length, within the range tested, does not significantly change compatibility.

Next, a systematic evaluation of the long-term efficacy of BP-Trz-FU-$C_{16}$ on inhibitory and virucidal efficacy was performed against HSV-1 and HCoV-229E. Here, multiple conditions were tested for up to five months under a range of temperatures (-20 °C, 4 °C and room temperature) and storage conditions (dry powder or suspended in phosphate-buffered saline (PBS) and water). Across all tested parameters, BPNS-Trz-FU-C16 showed remarkable preservation of activity. No significant loss in antiviral or virucidal efficacy was detected under any formulation or temperature tested.

To assess whether our virucidal BPNS-Trz-FU-$C_{16}$ could overcome viral mutations, HSV-1 was serially passaged for 13 generations in the presence of BPNS-Trz-FU-$C_{16}$ and, for comparison, under the selective pressure of acyclovir (ACV) or foscarnet. As expected, ACV-selected virus (P13) exhibited a pronounced loss of susceptibility, with a 14-fold increase in $IC_{50}$ relative to wild-type (WT), consistent with resistance arising from mutations in thymidine kinase or DNA polymerase (Fig. 6). In contrast, FOS-passaged virus revealed only a minor $IC_{50}$ shift, reflecting its direct polymerase inhibition mechanism and low dependence on viral enzyme activation. Remarkably, HSV-1 exposed to BPNS-Trz-FU-$C_{16}$ over the 13 passages did not develop measurable resistance. This high barrier to resistance is likely due to the viral deactivation of BPNS-Trz-FU-$C_{16}$, which combines viral surface binding and membrane disruption through long and hydrophobic alkyl chains. Single-target antivirals such as acyclovir (ACV) lose efficacy after single-point mutations. In contrast, BPNS-Trz-FU-C16 acts through synergistic electrostatic and hydrophobic interactions.

To determine whether prolonged exposure affected viral spread, plaque size distributions were analyzed after 13 passages. All treated viruses formed plaques comparable to WT, indicating that neither conventional antivirals nor BPNS-Trz-FU-C$_{16}$ significantly altered cell-to-cell spread. Although ACV-resistant variants emerged, these mutations did not compromise replication efficiency, consistent with previous observations that some ACV-resistant HSV-1 strains maintain robust *in vitro* growth. FOS-treated virus similarly retained normal plaque morphology, reflecting its low propensity for resistance and minimal impact on viral fitness. Collectively, these results demonstrate that while ACV rapidly drives resistance, BPNS-Trz-FU-C$_{16}$ imposes a substantially higher barrier. It thus defines a new class of materials against multiple viruses with low mutation susceptibility.

**Conclusion**

In summary, we report scalable, degradable, and biocompatible BP hybrids that exhibit mutation-resilient, antiviral activity against multiple viruses. Stepwise covalent functionalization of BP surfaces produces materials that both inhibit and irreversibly inactivate diverse viruses. These findings reveal synergistic electrostatic and hydrophobic interactions as the basis of durable virucidal performance. The work represents a major advance toward biodegradable 2D materials for broad-spectrum antiviral applications. It defines a new class of materials positioned between antiseptics and antiviral medicines.

**Acknowledgement**

This work is supported by the Federal Ministry of Research, Technology and Space (BMFTR) in the framework of NanoMatFutur (PathoBlock, project number 13XP5191). We would like to acknowledge the assistance of the Core Facility BioSupraMol supported by the DFG. We further thank for support from the SupraFAB research building realized with funds from the federal government and the city of Berlin. We would like to acknowledge Helmholtz Zentrum Berlin (HZB) team located at the BESSY II synchrotron radiation facility as well as Dr. M. Brzheszinskaya (HZB Berlin) from the HE-SGM collaborate research group. We would like to thank Marwin Raue for black phosphorus production support. We acknowledge Prof. Dr. Christian Müller for his administrative support regarding black phosphorus regulations within the university. The authors would like to thank Ana Hočevar, Priyadarshini Baidya, Lena Fredrich, Billy-Joe Buhrmeister, and Anastasiia Sydorova for the support within synthesis optimization as well as M. Eng. Jörg Stockmann from BAM 6.1 for his support with HAXPES data acquisition.

# Supplementary Information

# Scalable covalently functionalized black phosphorus hybrids for broad-spectrum virucidal activity

Na Xing[a#], Jasmin Er[ab#], Ricardo M. Vidal[c,d], Sandhya Khadka[a], Robert Schusterbauer[ab], Maik Rosentreter[a], Ranen Etouki[a], Rameez Ahmed[a], Taylor Page,[a] Philip Nickl[a], Obida Bawadkji[a], Anja Wiesner[a], Jörg Radnik[b], Vasile-Dan Hodoroaba[b], Kai Ludwig[e], Jakob Trimpert[de], Ievgen S. Donskyi[ab*]

[#] These authors contributed equally

[a]Institut für Chemie und Biochemie, Freie Universität Berlin, Takustraße 3, 14195 Berlin, Germany

[b]Federal Institute for Material Science and Testing (BAM), Division 6.1, Unter den Eichen 44-46, 12205 Berlin, Germany

[c]Institut für Virologie, Freie Universität Berlin, Robert-von-Ostertag-Straße 7-13, 14163 Berlin, Germany,

[d]College of Veterinary Medicine, Kansas State University, Mosier Hall, 1800 Denison Avenue, Manhattan, KS 66506-5600, USA.

[e]Forschungszentrum für Elektronenmikroskopie und Gerätezentrum BioSupraMol, Freie Universität Berlin, Fabeckstraße 36A, 14195 Berlin, Germany

*Corresponding author. E-mail: ievgen.donskyi@fu-berlin.de

**Table of contents**

# 1. Experimental Section

## 1.1. Materials and methods

### 1.1.1. Reaction conditions and chemicals

Unless otherwise stated all reactions were conducted under an inert argon atmosphere using schlenk techniques or in an inert glovebox. Red phosphorus (≥97.0%, SigmaAldrich), sodium azide (≥99.0%, NaN$_3$, Alfa Aesar), cyanuric chloride (≥99.0%, SigmaAldrich), tetrabutylammonium bromide (≥99.0%, TBAB, SigmaAldrich), O-(7-Azabenzotrial-1-yl)-*N,N,N',N'*-tetramethyluroniumhexafluorophosphate (≥99.0%, HATU, abcr), triethylamine (≥99.5%, Et$_3$N, SigmaAldrich), fucoidan (≥95.0%, *Undaria pinnatifida*, SigmaAldrich), epichlorohydrin (≥99.0%, SigmaAldrich), sodium hydroxide (≥98.0%, Fisher Scientific), ammonia solution (25%, SigmaAldrich), undecylamine (≥98.0%, TCI), tetradecylamine (≥96.0%, TCI), capric acid (≥98.0%, SigmaAldrich), chlorodiphenylphosphine (98.0%, Fisher Scientific), palmitic acid (≥99.0%, TCI), *N,N*-Diisopropylethylamine (≥99.0%, DIPEA, SigmaAldrich), triphenylphosphine (≥99.0%, SigmaAldrich), 3,5-dichloroaniline (≥98.0%, Fisher Scientific), aniline (≥99.9%, Acros organics) were purchased from commercial suppliers and used without further purification. Dry NMP, IPA and DMF were purchased from Acros organics and used as received unless otherwise stated. For all reactions that involved air- or water-sensitive compounds the solid reagents were dried on the schlenk line (1·10$^{-3}$ mbar) at least one day prior use. Water was used from Mili-Q® Advantage A10 Water Purification System. Purification was performed in Spectra/Por® 6 cellulose dialysis tubes (100 kDa, 50 kDa and 2 kDa MWCO, Repligen).

### 1.1.2. High Energy Planetary Ball-Mill (HEPBM)

The Planetary Mono Mill Pulverisette 6 *classic line* from Fritsch GmbH (Germany) was used for the conversion process. The 80 mL agate grinding bowl (Fritsch GmbH, Germany) made of hardened stainless steel was only opened inside the glovebox and cleaned after every use using dry IPA. Stainless steel balls with a diameter of 10 mm were used (Fritsch GmbH, Germany). Milling time, pause time and number of repetitions was varied. All experiments were performed at an rpm of 600.

### 1.1.3. Nuclear magnetic resonance (NMR) spectroscopy

1H-NMR measurements were performed on a Joel ECZ600 (Japan) and a Bruker AVANCE III 500 (USA). Processing of the data was done using the software MestReNova (version 14.3.0). Chemical shifts δ were given in ppm in relativity to an internal standard, usually the deuterated solvent D2O (δ (1H) = 4.79 ppm) or CDCl3 (δ (1H) = 7.26 ppm).

### 1.1.4. Raman spectroscopy

Raman spectroscopy was conducted using a Horiba XploRA Plus™ spectrometer (Japan), equipped with a 532 nm Nd:YAG laser and a motorized x/y piezo stage. The laser beam was focused through a 100x Nikon® objective lens, with prepared samples being irradiated at 1 mW (1% of 100 mW, with the actual energy reduced by a filter). The spectrometer, which was autocalibrated with a silicon wafer prior to measurements, was set to a 2400 $cm^{-1}$ grating centered at 600 $cm^{-1}$. The samples were deposited on 1 cm x 1 cm silicon wafers. For statistical raman spectroscopy, raman maps were generated from selected sample regions using the point by point method, with substrate movement during mapping controlled by the motorized x/y piezo stage. Average spectra were created from these maps. To analyze the data, a polynomial baseline fit was applied in LabSpec 6 (HORIBA).

### 1.1.5. Infrared (IR) spectroscopy

IR spectra were recorded using a Spectrum Two FT-IR spectrometer by PerkinElmer (USA). The measuring range was set between 4000 cm$^{-1}$ to 450 cm$^{-1}$. Samples were measured at room temperature by adding small amounts of sample to the crystal.

### 1.1.6. X-ray photoelectron spectroscopy (XPS)

XPS measurements were performed using an EnviroESCA spectrometer (SPECS Surface Nano Analysis GmbH, Berlin, Germany), equipped with a monochromatic Al Kα X-ray source (Excitation Energy = 1486.71 eV) and a PHOIBOS 150 electron energy analyzer operating in fixed analyzer transmission (FAT) mode. All spectra were acquired under ultra-high vacuum. Samples for XPS analysis were prepared on silicon wafers or indium foil. The spectra were measured in normal emission, and a source-to-sample angle of 60° was used. Instrument calibration followed the technical procedure provided by SPECS (calibration was performed according to ISO 15472). Survey spectra were acquired with a pass energy of 80 electron volt (eV), and the high resolution XP spectra were acquired with a pass energy of 50 eV. Raw data fitting was performed using the UNIFIT 2022 software. For fitting, a Shirley background and a Lorentzian-Gaussian (L-G) sum function were used. Unless stated otherwise, a L-G mixing ratio of 0.3 was used for carbon peaks and 0.40 for heteroatom peaks. If not indicated otherwise, all binding energies were referenced to the binding energy of sp$^3$-hybridized C–C bond component at 285 eV or the energy of C-N=C at 285.5 eV.

### 1.1.7. Hard X-ray photoelectron spectroscopy (HAXPES)

HAXPES experiments were carried out using a monochromatic Cr Kα source (5.4 keV, Ulvac-PHI, Chanhassan, USA). For charge neutralization, low-energy electrons and Ar$^+$-ions were employed. The utilized spot size was 100 µm and photoelectrons were detected at an emission angle of 45° relative to the surface. Measurements were performed at pressures between 10$^{-8}$ and 10$^{-10}$ mbar, with samples mounted to the sample holder using adhesive tape. High-resolution HAXPES spectra were measured using a pass energy of 55 eV, a step size of 0.1 eV and 10-20 sweeps, depending on the signal-to-noise ratio of each element.

### 1.1.8. Near edge X-ray adsorption fine structure (NEXAFS)

NEXAFS experiments were carried out at the synchrotron radiation source BESSY II (Berlin, Germany) at the HE-SGM monochromator dipole magnet CRG beamline. NEXAFS spectra were acquired in total energy electron yield (TEY) mode using a channel plate detector. The resolution E/ΔE of the monochromator at the carbonyl $\pi^*$ resonance (hv = 287.4 eV) was in the order of 2500. Raw 3 spectra were divided by ring current and monochromator transmission, the latter obtained with a freshly sputtered Au sample. Alignment of the energy scale was achieved by using an $I_0$ feature referenced to a C1s → $\pi^*$ resonance at 285.4 eV measured with a fresh surface of HOPG (highly ordered pyrolytic graphite, Advanced Ceramic Corp., Cleveland, USA). If not otherwise denoted, all NEXAFS spectra are shown after subtraction of the pre-edges followed by normalization of the post-edge count rates to one. All C K-edges were measured at 55° incident angle of the linearly polarized synchrotron light.

### 1.1.9. Time-of-flight secondary mass spectrometry (ToF-SIMS)

ToF-SIMS analysis was performed using a TOF-SIMS M6 instrument (IONTOF GmbH, Münster, Germany). The samples were analyzed at room temperature. Measurements were conducted in collimated burst alignment mode with a 30 keV $Bi_3^+$ primary ion beam in positive and negative polarity. A field of view (FoV) measuring 150 × 150 μm was rastered in random mode with 256 × 256 pixels, acquiring one shot per pixel. Spectra were calibrated using typical organic fragments ($CH_2^+$ at 14.02, $C_2H_4^+$ at 28.03, $C_3H_6^+$ at 42.05, $C_4H_8^+$ at 56.06 for positive and $C^-$ at 12.00, $C_2^-$ at 24.00, $C_3^-$ at 36.00, $C_4^-$ at 48.00 m/z for negative mode).

### 1.1.10. Zeta-potential

Zeta-potential measurements were carried out in Milli-Q water at ambient conditions using a Zetasizer Ultra (Malvern Panalytical, UK). Samples were analyzed in a folded capillary zeta cell (Malvern Panalytical, UK) operated in automatic mode. Each measurement was repeated three times and the average zeta-potential value was reported.

### 1.1.11. Thermogravimetric analysis (TGA)

TGA experiments were performed on a TGA 8000 (PerkinElmer, USA) under a nitrogen atmosphere with a nitrogen flow of 40 ml/min. Samples containing black phosphorus were purged with nitrogen prior measurement for 1 h. The heating rate was set to 10 °C/min and the instrument was calibrated using calcium oxalate. $Al_2O_3$ crucibles were used and the sample mass varied from 1 to 3 mg.

### 1.1.12. Atomic force microscopy (AFM)

All samples were imaged with a JPK nanowizard AFM (Bruker) with in AC mode (also known as tapping mode). All experiments were performed in air and samples were prepared as follows. Rectangular pieces of muscovite mica or silicon wafers of about 1 $cm^2$ were used as a substrate. Double sided tape was used to glue the mica to circular metal pucks. It was then cleaved with regular adhesive tape to obtain a clean flat surface. 10 µl of the samples dispersed in water or MeOH (Concentration: 0.1 $mgml^{-1}$) was placed in the middle and allowed to spread and eventually dry on the surface. The samples were then mounted on the XY stage. TAP300Al-G silicon AFM probes from Budget sensors with a nominal spring constant of 40 N $m^{-1}$ and tip radius of less than 10 nm were used. Scan rates were usually 1 Hz with 256 points per line. The JPKSPM data processing software Version 6.1.74 (trademark JPK instruments) was used for Image analysis. In images obtained with sheets, a XY plane fit order 1 was performed, followed by a flatten order 1 as well. Determination of sheet thickness was done using the cross-section analysis function.

### 1.1.13. Cryo-Transmission electron microscopy (Cyro-TEM)

HCoV-229E were purified and concentrated by ultracentrifugation with 20 % sucrose solution. Purified virions were incubated with BPNS-Trz-FU-$C_{16}$ for 1 h at 37 °C. Fixation was conducted using 1% glutaraldehyde for 1 h. 5 µL of aqueous PBS dispersion of BPNS-Trz-FU-$C_{16}$ (5 mg/mL) and purified HCoV 229E were placed onto glow-discharged, perforated, carbon film-covered microscopical 200 mesh grids (R1/4 batch of Quantifoil MicroTools GmbH, Jena, Germany) and vitrified by automatic blotting and plunge freezing with a FEI Vitrobot Mark IV (Thermo Fisher Scientific Inc., Waltham, Massachusetts, USA) using liquid ethane as cryogen. The vitrified specimens were transferred to a FEI TALOS ARCTICA electron microscope (Thermo

Fisher Scientific Inc., Waltham, Massachusetts, USA). Micrographs were acquired on a FEI Falcon 3 direct electron detector.

### 1.1.14. Cell culture and virus strains

Vero E6 cells (ATCC CCL-81) were cultured in Dulbecco's modified Eagle's medium (DMEM) supplemented with 10% fetal bovine serum (FBS), 100 U/mL penicillin, and 100 μg/mL streptomycin. Crandell-Rees Feline Kidney (CRFK) cells and Huh-7 cells, a line of human liver cancer cells, derived from a hepatocellular carcinoma cells were grown in DMEM containing 10% FBS, 100 U/mL penicillin, and 100 μg/mL streptomycin. All cells were cultured at 37°C and 5% $CO_2$.

The HSV-1 strain F carrying GFP (kindly provided by Y. Kawaguchi, University of Tokyo, Japan) was propagated on Vero E6 cells. The feline coronavirus (FCoV) stain WSU 79-1683 and human coronavirus strain 229E were propagated on CRFK and Huh-7 cells, respectively.

### 1.1.15. Cell viability assays

To analyze the cytotoxicity and determine the half-maximal cytotoxic concentration (CC50) of compounds, the CellTiter-Glo® Luminescent Cell Viability Assay Kit was used according to the manufacturer's instructions. Briefly, cells were seeded in an opaque-walled 96-well plate and cultured in a $CO_2$ incubator at 37°C for 24 h. After overnight incubation, various concentrations of compounds were added to the cells and incubated for another day Then, equal amount of CellTiter-Glo® Reagent was added to the cell culture medium and mixed for 2 minutes on an orbital shaker to induce cell lysis. At room temperature the plate was incubated for 10 minutes to stabilize the luminescent signal. Subsequently, the luminescence was determined by a microplate reader.

### 1.1.16. Plaque reduction assay

10-fold serially diluted samples were incubated with 100 PFU of viruses for 45 min at 37°C. Then, the viruses and compounds mixture were added on a confluent monolayer of cells. Virus adsorption was allowed for 1 h before the supernatant was aspirated and replaced by a semi-solid overlay. After 48 h or 72 h, the overlay was removed, and cells were fixed with 4% formalin and stained with crystal violet. The percentage of virus inhibition was calculated by the plaque numbers of virus control and samples.

### 1.1.17. Pre-infection inhibition assay

Cells grown in a 96-well plate at 100% confluence were treated with various concentrations of antiviral compounds (ranging from 0.1 μg/mL to 1 mg/mL) for 45 min at 37 °C with continuously shaking and then infected by viruses at a MOI of 0.01 for 48 h. Afterwards, HSV-1 infected cells were stained with Hoechst 33342 and fixed with 4% formaldehyde in phosphate-buffered saline (PBS) for 30 min. For 229E or FCoV pre-infection assay, infected cells were fixed and permeabilized with 0.1% Triton X-100 in PBS for 10 min, and blocked with 3% bovine serum albumin (BSA) in PBS for 30 min. Cells were then incubated with primary antibodies (mouse anti-IBV monoclonal antibody provided by Gary R. Whittaker or Rabbit anti-229E polyclonal antibody from Bio-Rad) for 1 h, washed three times with PBS, and stained with secondary antibodies anti-mouse Alexa Fluor 488 fluorophores or anti-rabbit Alexa Fluor 594 fluorophores (Thermo Fisher Scientific) for 1 h with gentle shaking. The cell nucleus was stained with Hoechst 33342. Images were acquired using Zeiss fluorescent microscope and the percentage of infection inhibition were analyzed.

## 1.2. Synthesis

### 1.2.1. Screening experiment to produce black phosphorus (BP)

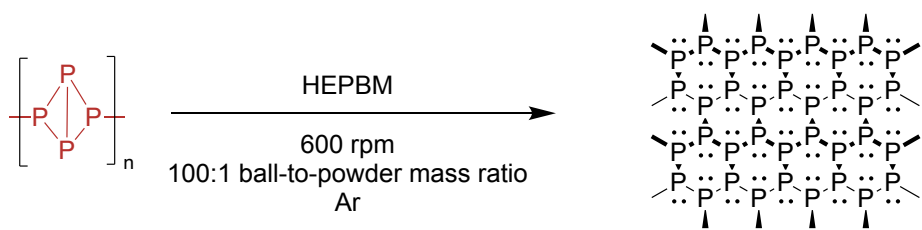

**Scheme S1.** Scheme for the production of BP.

Inside a glovebox, dried RP (1.00 g) was transferred into an 80 mL agate milling chamber containing 25 steel balls (~100 g). The chamber was sealed under an argon atmosphere using a screw clamp system to maintain the inert environment and subsequently placed on the Pulverisette 6. Milling was conducted at a rotational speed of 600 rpm under various milling duration, pause intervals and cycle repetitions (table S1). In runs, where pause times where applied, the rotation direction was reversedafter each pause to promote more uniform conversion. Upon completion, the resulting BP was collected inside the glovebox.

**Table S1.** Ball milling parameters for the synthesis the synthesis of BP, including milling time, pause times, number of cylcles, reverse mode and total milling time.

| Milling time per Cycle [min] | Pause time [min] | Cycles | Reverse mode after pause | Total milling time [min] |
|---|---|---|---|---|
| 1 | - | 1 | No | 1 |
| 5 | - | 1 | No | 5 |
| 15 | - | 1 | No | 15 |
| 15 | 15 | 2 | Yes | 30 |
| 30 | - | 1 | No | 30 |
| 30 | 15 | 1 | Yes | 60 |
| 30 | 30 | 4 | Yes | 120 |
| 30 | 30 | 8 | Yes | 240 |

### 1.2.2. Liquid phase exfoliation (LPE) of BP with an ultrasonic bath

Inside a glovebox, bulk BP (1.00 g) was placed inside a 250 mL round bottom flask and sealed before removal from the argon atmosphere. NMP (100 mL) was added to reach a concentration of 10 mg/mL. The suspension was sonicated in a bath sonicator operating at full power for 8 h at room temperature. After, the dispersion was divided between two centrifuge tubes and subjected to centrifugation (45 min, 1500 rpm, 4 °C) to remove bulky material. The upper 2/3 were collected. To purify the BPNS, the supernatant was further centrifuged (6 x 15 min, 9500 rpm, 4 °C) and redispersed in acetone between each cycle. The final product was obtained as a black powder after lyophilization (~10 mg) and stored under argon atmosphere.

### 1.2.3. LPE of BP with probe sonication

Inside a glovebox, bulk BP (1.00 g) was placed inside a 100 mL round bottom flask and sealed before removal from the argon atmosphere. NMP (100 mL) or degassed water (100 mL) was added to achieve a concentration of 10 mg/mL. The suspension was sonicated using a tip sonicator submerged directly into the dispersion operated at full power for 1 h under ice cooling. The resulting dispersion was transferred into two centrifuge tubes and subjected to centrifugation (45 min, 1500 rpm, 4 °C) to remove bulky material. The upper 2/3 were collected. To purify the BPNS exfoliated in NMP, the suspension was further centrifuged (6 x 15 min, 9500 rpm, 4 °C) and redispersed in acetone between each cycle. Finally, the material was lyophilized. BPNS exfoliated in water were lyophilized directly. The final products were obtained as black powders (~100 mg (NMP); ~80 mg ($H_2O$)) and stored under argon atmosphere.

### 1.2.4. Synthesis of BPNS-Trz

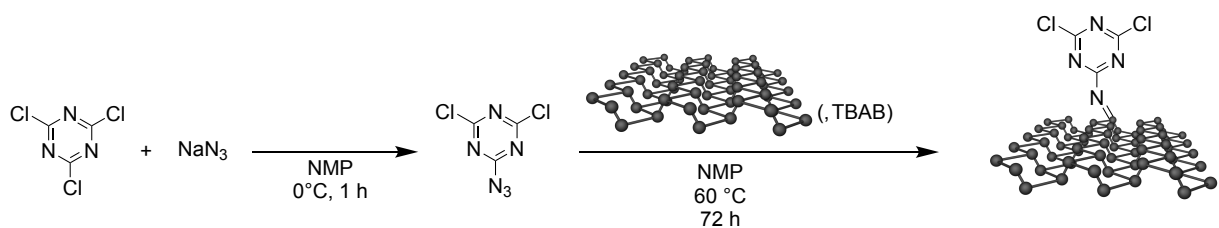

**Scheme S2.** Synthetic pathway of BP-Trz.

Cyanuric chloride (2.40 g, 13.0 mmol, 1.0 eq.) was dissolved in dry NMP (20 mL) in a 100 mL schlenk flask and cooled to 0 °C in an ice bath. Sodium azide (0.84 g, 12.9 mmol, 1.0 eq.) was added slowly upon stirring in an argon counterflow. The reaction mixture was stirred for at 0°C for 1 h to form cyanuric azide. In parallel, exfoliated BPNS were redispersed in NMP (40 mL) and a catalytic amount of TBAB was added. The dispersion was purged with nitrogen for 20 min and added to the cyanuric azide solution. The reaction mixture was stirred for 72 h at 60 °C. After cooling to room temperature, the product was isolated by centrifugation (20 min, 9500 rpm, 4 °C). Further, the residue was washed by six repeated washing steps using a 1:1 mixture of acetone and deionized water (50 mL). The product was obtained as black powder after lyophilization (150 mg (NMP); 50 mg ($H_2O$)) and stored under argon.

### 1.2.5. Synthesis of FU-NH₂

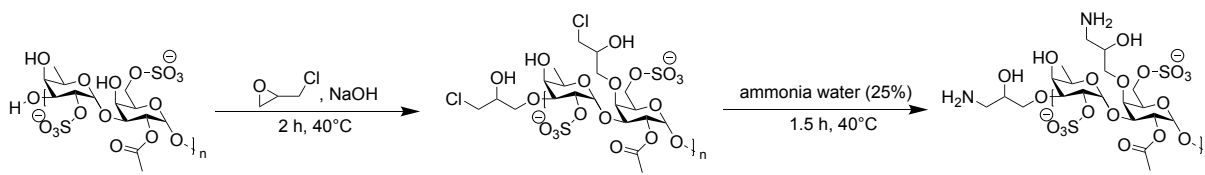

**Scheme S3.** Synthetic pathway of FU-NH₂.

FU (0.20 g) from *Undaria pinnatifida* was dissolved in water (7.0 mL). Under stirring NaOH (2.6 M, 2.3 mL, 6.0 mmol, 0.36 eq.) and epichlorohydrin (1.4 mL, 17.8 mmol, 1 eq.) were added. After stirring for 2 h at 40°C the reaction solution was dialyzed against water for 3 d (2 kDa MWCO). The lyophilized product (0.19 g) was redissolved in aqueous ammonia solution (25%, 1.5 mL) and stirred for at 40°C for 90 min. For purification, the crude product was dialyzed against water for 3 d (2 kDa MWCO). After lyophilization, the product (0.18 g, 90%) was obtained as a white powder.

### 1.2.6. Synthesis of BPNS-Trz-FU

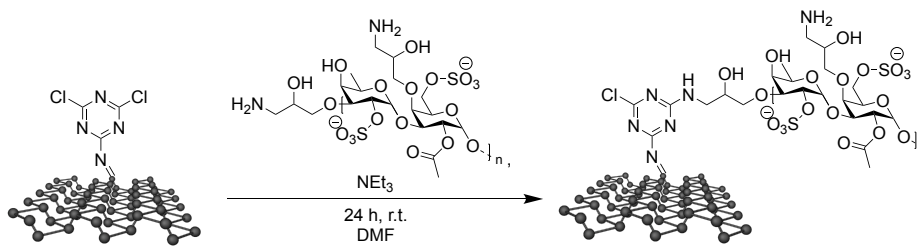

**Scheme S4.** Synthetic pathway of BPNS-Trz-FU.

FU-NH₂ (0.10 g) was dissolved in dry DMF (15 mL) and NEt₃ (0.7 mL, 5.0 mmol) was added dropwise under stirring. BPNS-Trz (0.02 g) was added to the mixture and the suspension was sonicated for 1 h. The reaction was then stirred at room temperature for 48 h. Following completion, the crude product was dialyzed against water for 3 d (100 kDa MWCO). After lyophilization the product (74.2 mg (NMP); 80.2 mg (H₂O)) was obtained as black powder and stored under argon.

### 1.2.7. Synthesis of BPNS-Trz-FU/C$_x$

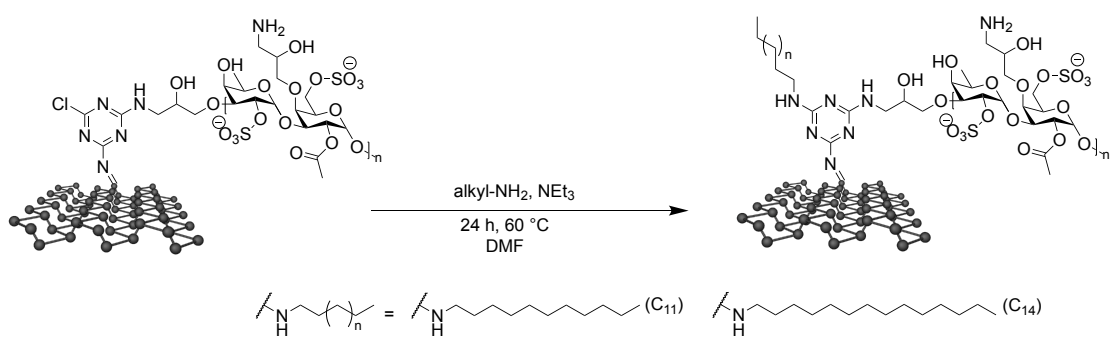

**Scheme S5.** Synthetic pathway of BPNS-Trz-FU/C$_x$.

In a 100 mL Schlenk-flask, the corresponding primary alkyl amines (0.10 g, 1.0 eq., 5:1 mass ratio to BPNS-Trz-FU) were dissolved in DMF (5 mL) and BPNS-Trz-FU (0.02 g) was added. The solution was then sonicated at room temperature for 30 min. Subsequently, triethylamine (1 eq.) was added to the reaction and the mixture was heated to 60°C and was stirred for 48 h. After the reaction, the crude product was purified by dialysis against MeOH for 3 d (100 kDa MWCO). After lyophilization, the product (BPNS-Trz-FU/C$_{11}$ (NMP): 18.9 mg; BPNS-Trz-FU-C$_{14}$ (NMP): 18.6 mg) was obtained as a grey-black powder and stored under argon.

### 1.2.8. Synthesis of BPNS-Trz-FU-C$_x$

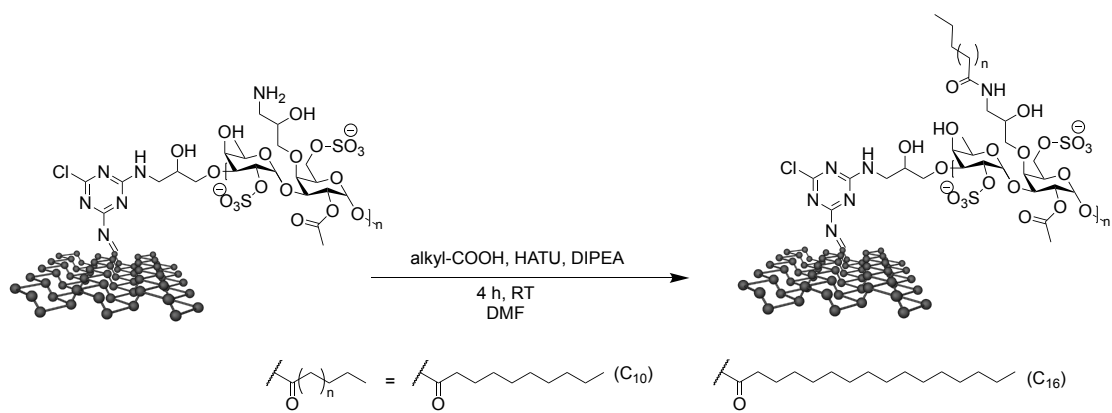

**Scheme S6.** Synthetic pathway of BPNS-Trz-FU-C$_x$.

In a 100 mL Schlenk-flask, the corresponding fatty acid (0.15 g, 1.0 eq., 5:1 mass ratio to BPNS-Trz-FU), HATU (1.0 eq.), and DIPEA (2.0 eq.), were dissolved in DMF (5 mL) and placed in an ultrasonic bath for 5 min. The solution was then stirred at room temperature for 10 min, before adding BPNS-Trz-FU (0.03 g). Subsequently, the rection mixture was stirred at room temperature for 4 h. After the reaction, the crude product was purified by dialysis against MeOH for 3 d (100 kDa MWCO). After lyophilization, the product (BPNS-Trz-FU-$C_{10}$ (NMP): 28.6 mg; BPNS-Trz-FU-$C_{10}$ ($H_2O$): 30.6 mg; BPNS-Trz-FU-$C_{16}$ (NMP): 38.8 mg; BPNS-Trz-FU-$C_{16}$ ($H_2O$): 32.2 mg) was obtained as a grey-black powder and stored under argon.

### 1.2.9. Synthesis of 2-azido-4,6-dichloro-1,3,5-triazine

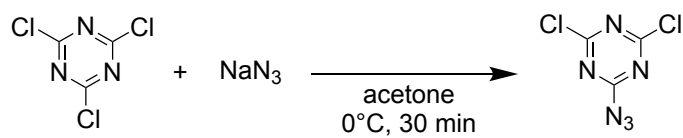

**Scheme S7.** Scheme for the synthesis of 2-azido-4,6-dichloro-1,3,5-triazine.

Following a protocol from Sharma et al.[1] 2-azido-4,6-dichloro-1,3,5-triazine was synthesized. Briefly, cyanuric chloride (5.0 g, 27.3 mmol, 1.0 eq.) was dissolved in acetone (100 mL) and cooled to 0°C in an ice bath. In parallel, sodium azide (1.77 g, 27.3 mmol, 1.0 eq.) was dissolved in water (50 mL) and cooled to 0°C. The aqueous sodium azide solution was added dropwise to the cyanuric chloride solution under vigorous stirring while maintaining 0°C. The mixture was stirred at this temperature for until completion (30 min), before removing the acetone under reduced pressure at ice cold conditions. The remaining aqueous phase was extracted with cold dichloromethane (3 x 50 mL), the combined organic layers were dried over $MgSO_4$ and filtered. The solvent was removed under reduced pressure at ice cold conditions. The crude product was purified using column chromatography (*n*-hexane). The product (4.3 g, 83%) was obtained as a colorless solid.

### 1.2.10. Synthesis of 3,5-dichloro-*N*-(triphenylphosphoranylidene)aniline

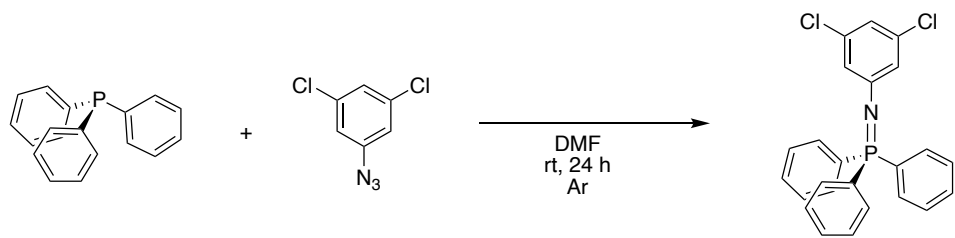

**Scheme S8.** Scheme for the synthesis of 3,5-dichloro-*N*-(triphenylphosphoranylidene)aniline.

Following a modified protocol from Meguro et al.[2] 3,5-dichloro-*N*-(triphenylphosphoranylidene)aniline was synthesized. In a 10 mL schlenk-flask 1-azido-3,5-dichlorobenzene (0.05 g, 0.31 mmol, 1.0 eq.) was dissolved in dry DMF (2 mL). Triphenylphosphine (0.097 g, 0.37 mmol, 1.2 eq.) was added in an argon counter flow under stirring. The reaction mixture was stirred at room temperature for 24 h. After completion, the solvent was removed under reduced pressure. The crude product was purified using column chromatography (*n*-pentane/EtOAc = 5/1) to give the product (0.05 g, 38%) as a colorless solid.

**¹H-NMR** (600 MHz, CDCl$_3$): $\delta$ [ppm] = 7.75-7.69 (m, 6H), 7.58-7.54 (m, 3H), 7.50-7.46 (m, 6H), 6.62-6.60 (m, 3H).

**³¹P-NMR** (600 MHz, CDCl$_3$): $\delta$ [ppm] = 5.72 (m).

**MS** (ESI): m/z = 422.06 [M]$^{+\cdot}$.

### 1.2.11. Synthesis of *N*-(4,6-dichloro-1,3,5-triazine)triphenylphosphoranylidene

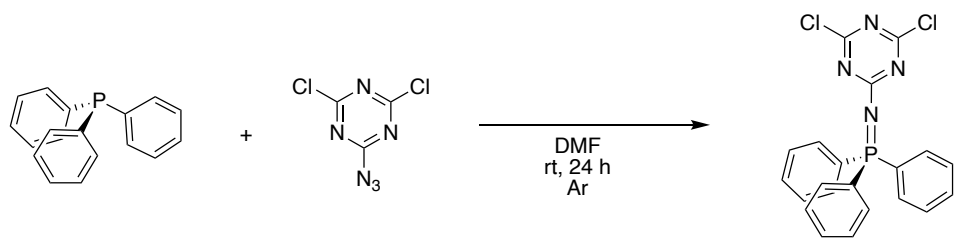

**Scheme S9.** Scheme for the synthesis of *N*-(4,6-dichloro-1,3,5-triazine)triphenylphosphoranylidene.

Following a modified protocol from Meguro et al.[2] *N*-(4,6-dichloro-1,3,5-triazine)triphenylphosphoranylidene was synthesized. In a 10 mL schlenk-flask 2-azido-4,6-dichloro-1,3,5-triazine (0.10 g, 0.52 mmol, 1.0 eq.) was dissolved in dry DMF (3 mL). Triphenylphosphine (0.165 g, 0.63 mmol, 1.2 eq.) was added in an argon counter flow under stirring. The reaction mixture was stirred at room temperature for 24 h. After completion, the solvent was removed under reduced pressure. The crude product was purified using column chromatography (*n*-pentane/EtOAc = 5/1) to give the product (0.12 g, 54%) as a colorless solid.

**¹H-NMR** (600 MHz, CDCl$_3$): $\delta$ [ppm] = 7.84-7.78 (m, 2H), 7.63-7.58 (m, 1H), 7.52-7.48 (m, 2H).

**³¹P-NMR** (600 MHz, CDCl$_3$): $\delta$ [ppm] = 22.48 (m).

**MS** (ESI): m/z = 447.03 [M-Na]$^+$, 463.00 [M-K]$^+$.

### 1.2.12. Synthesis of N-(diphenylphosphaneyl)-N-1,1-triphenylphosphanamine

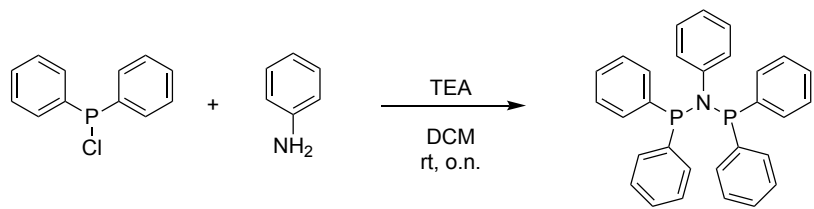

**Scheme S10.** Scheme for the synthesis of N-(diphenylphosphaneyl)-N-1,1-triphenylphosphanamine.

Following a protocol from Eady et al.[3] N-(diphenylphosphaneyl)-N-1,1-triphenylphosphanamine was synthesized. In a 10 mL schlenk-flask chlorodiphenylphosphine (1.08 mL, 2.0 mmol, 2 eq.) was added dropwise to a solution of aniline (0.09 mL, 1.0 mmol, 1 eq.) and triethylamine (1.4 mL, 10 mmol, 10 eq.) in DCM (5 mL) under stirring. The formation of a white precipitated could be observed. The reaction mixture was stirred at room temperature for 12 h. After completion, the solvent was removed under reduced pressure and the solid was thoroughly washed with MeOH (5 x 20 mL). The crude product was recrystallized from dichloromethane/hexane (1:1) at room temperature. The product (0.036 g, 78%) was obtained as a colorless solid.

**$^1$H-NMR** (600 MHz, CDCl$_3$): δ [ppm] = 7.38-7.29 (m, 20H), 6.98-6.92 (m, 3H), 6.67-6.64 (m, 2H).
**$^{31}$P-NMR** (600 MHz, CDCl$_3$): δ [ppm] = 69.12 (s).
**MS** (ESI): m/z = 462.15 [M-H]$^+$.

## 2. Supplementary figures and tables

**Table S2.** Characterization of BP derivatives starting from pristine BP, with elemental analysis and zeta potential values taken at 1.0 mg mL$^{-1}$.

| Compound | ζ [mV] | C [%] | H [%] | N [%] | S [%] |
|---|---|---|---|---|---|
| BPNS | -30.7 | 2.7 | 2.0 | 0.0 | 0.0 |
| BPNS-Trz | -33.6 | 17.7 | 4.5 | 5.7 | 0.0 |
| Fuc-NH$_2$ | -42.3 | 25.3 | 5.0 | 2.1 | 9.3 |
| BPNS-Trz-FU | -40.5 | 32.7 | 5.4 | 1.1 | 5.8 |
| C$_{11}$-NH$_2$ | 32.8 | 78.4 | 16.4 | 7.1 | 0.0 |
| C$_{14}$-NH$_2$ | 37.7 | 84.0 | 19.0 | 6.5 | 0.0 |
| BPNS-Trz-FU/C$_{11}$ | -30.1 | 59.3 | 10.3 | 4.1 | 2.1 |
| BPNS-Trz-FU/C$_{14}$ | -22.3 | 68.6 | 14.4 | 4.3 | 1.1 |
| C$_{10}$-OOH | -10.4 | 62.3 | 18.1 | 1.2 | 0.0 |
| C$_{16}$-OOH | -7.5 | 73.6 | 19.4 | 1.3 | 0.0 |
| BPNS-Trz-FU-C$_{10}$ | -42.3 | 53.5 | 12.2 | 2.6 | 5.7 |
| BPNS-Trz-FU-C$_{16}$ | -35.2 | 68.7 | 15.0 | 3.1 | 5.3 |

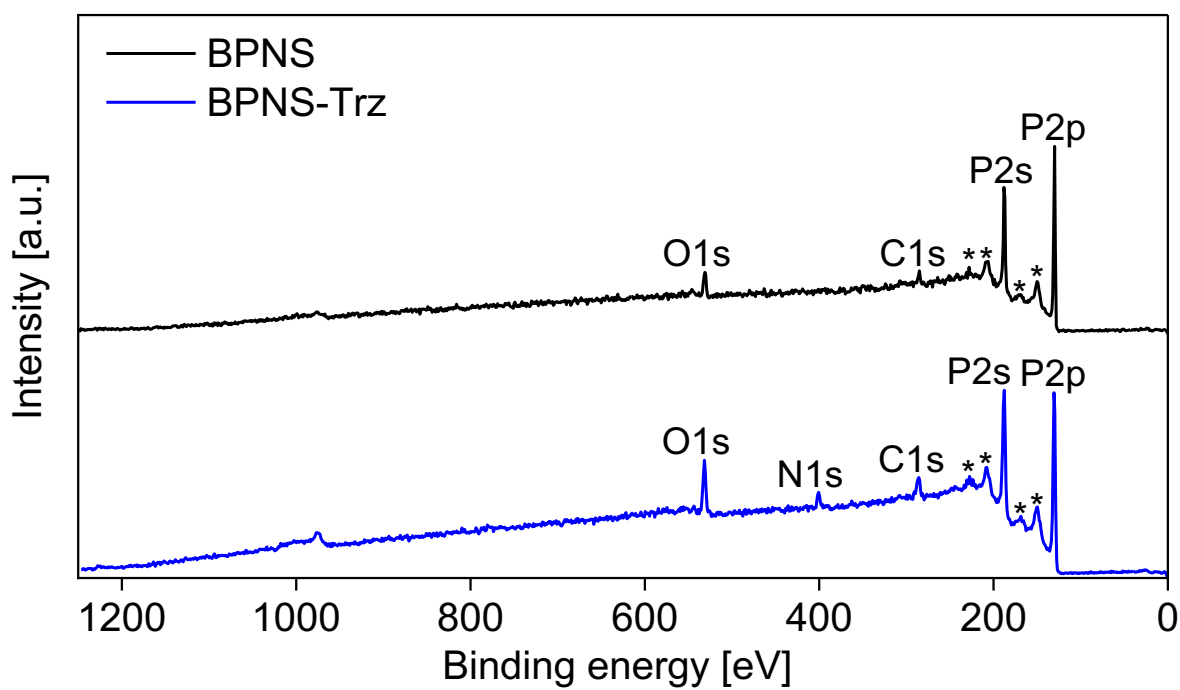

**Figure S1.** XPS survey spectra for BPNS and BPNS-Trz (peaks marked with * correspond to signals from the indium substrate background).

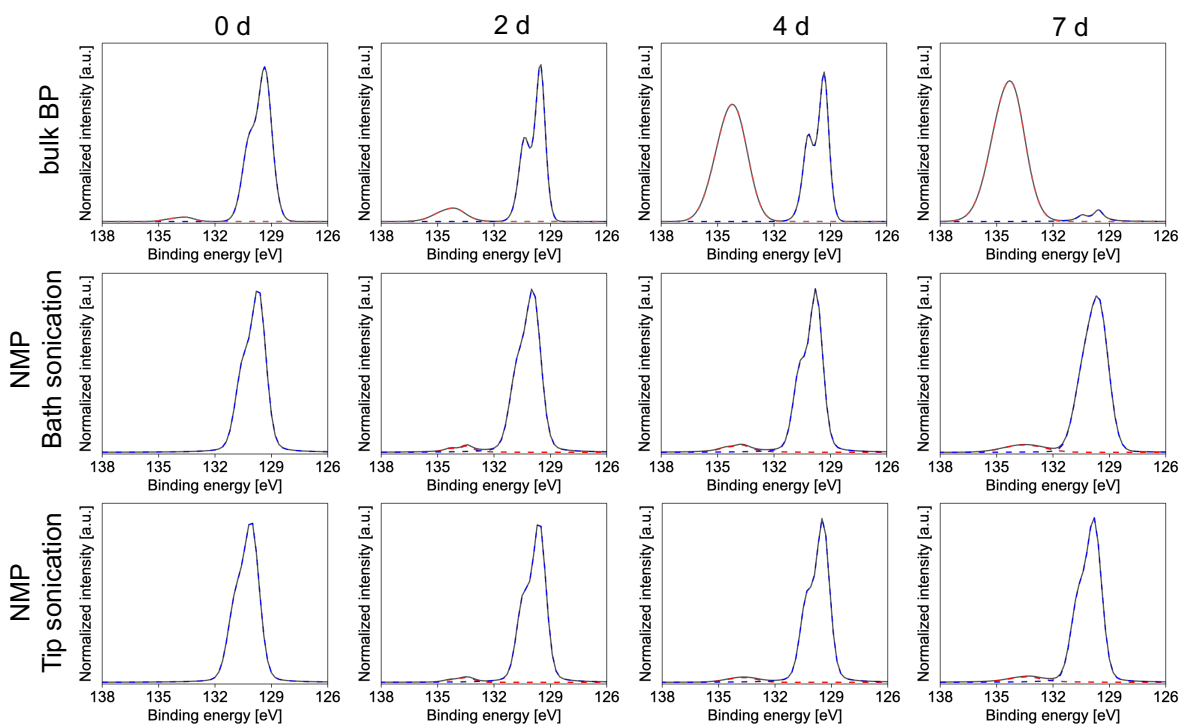

**Figure S2.** Stability experiments for the degradation of BP and BPNS. High resolution XPS P2p spectra for bulk BP, BPNS exfoliated via bath sonication in NMP and BPNS exfoliated via probe sonication in NMP after 0d, 2d, 4d and 7d.

**Table 3.** Relative elemental fractions from the quantification of the XPS survey of BPNS and BPNS-Trz.

|  | BE | BPNS at% | BPNS-Trz at% |
|---|---|---|---|
| O1s | 530 eV | 11% | 18% |
| N1s | 400 eV | 0% | 7% |
| C1s | 285 eV | 7% | 12% |
| P2s | 187 eV | 38% | 28% |
| P2p | 130 eV | 46% | 37% |

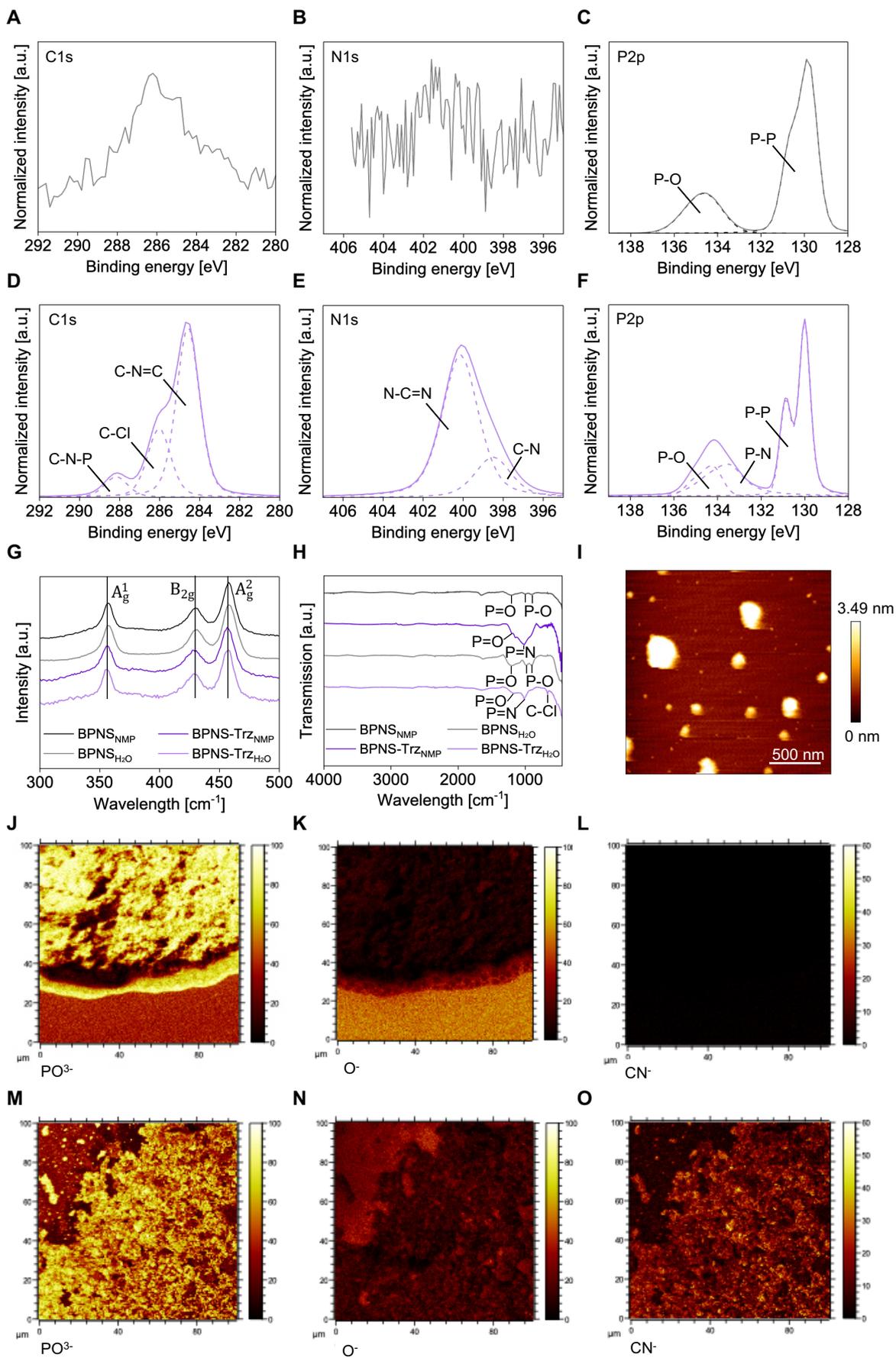

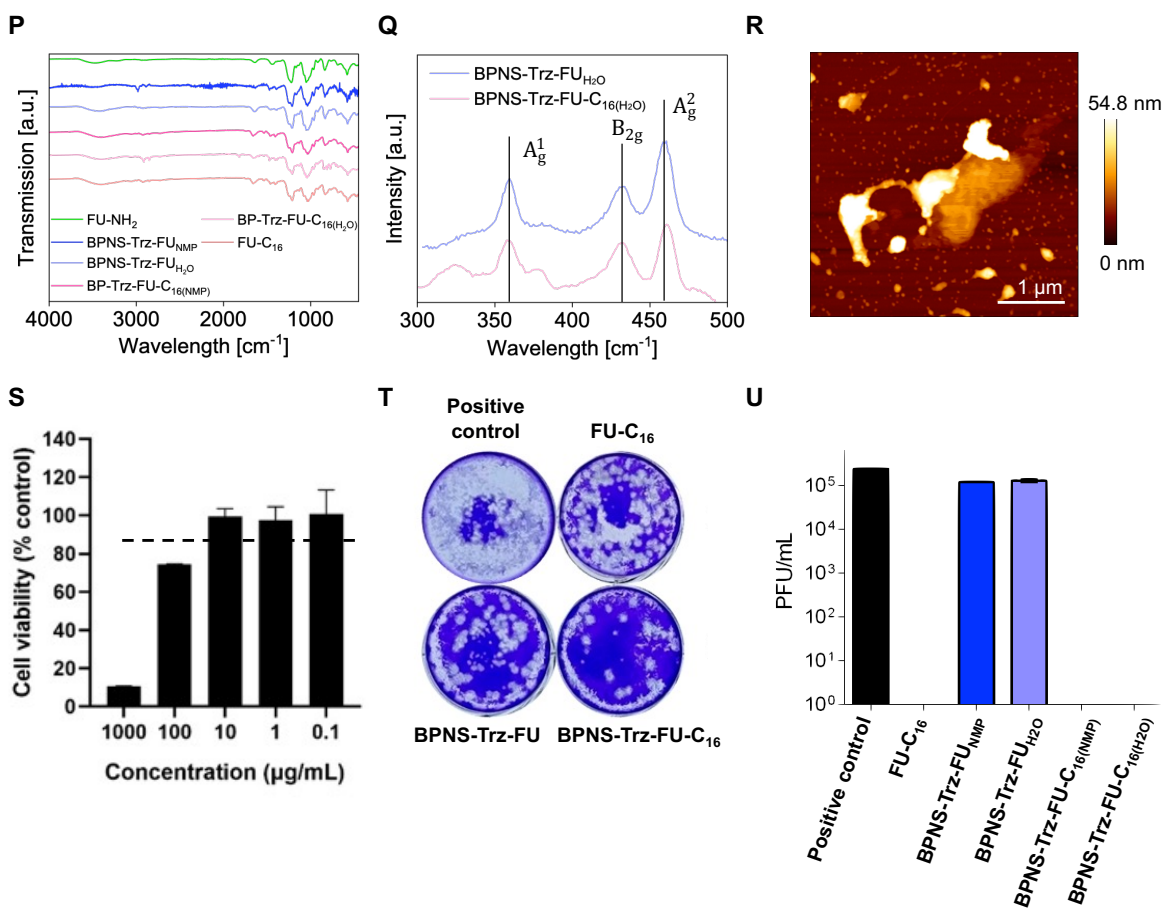

**Figure S3.** Characterization of BPNS derivatives exfoliated in water (BPNS$_{H2O}$). High resolution XPS spectra with peak fitting for BPNS$_{H2O}$ (A) C1s, (B) N1s and (C) P2p. High resolution XPS spectra with peak fitting for BPNS-Trz$_{H2O}$ (D) C1s, (E) N1s and (F) P2p. (G) Raman comparison of BPNS$_{NMP}$, BPNS$_{H2O}$, BPNS-Trz$_{NMP}$ and BPNS-Trz$_{H2O}$. (H) FTIR comparison of BPNS$_{NMP}$, BPNS$_{H2O}$, BPNS-Trz$_{NMP}$ and BPNS-Trz$_{H2O}$. (I) AFM of BPNS$_{H2O}$ (the scalebar corresponds to 500 nm). ToF-SIMS in the negative mode of BPNS$_{H2O}$ fragments (J) PO$_3^-$, (K) O$^-$ and (L) CN$^-$. ToF-SIMS in the negative mode of BPNS-Trz$_{H2O}$ fragments (M) PO$_3^-$, (N) O$^-$ and (O) CN$^-$. SEM-EDX mapping of the surface of BPNS-Trz (the scale bar corresponds to 10 μm). (P) FTIR comparison of FU-NH$_2$, BPNS-Trz-FU$_{NMP}$, BPNS-Trz-FU$_{H2O}$, BPNS-Trz-FU-C$_{16(NMP)}$, BPNS-Trz-FU-C$_{16(H2O)}$ and FU-C$_{16}$. (Q) Raman of BPNS-Trz-FU$_{H2O}$ and BPNS-Trz-FU-C$_{16(H2O)}$. (R) AFM of BPNS$_{H2O}$ (the scalebar corresponds to 1 μm). (S) Result of the cytotoxicity assay for BPNS-Trz-FU-C$_{16(H2O)}$. (T) Plaque reduction assay of different materials at 1 mM against HSV-1. (U) Result of the virucidal assay against HSV-1 for FU-C$_{16}$, BPNS-Trz-FU$_{NMP}$, BPNS-Trz-FU$_{H2O}$, BPNS-Trz-FU-C$_{16(NMP)}$ and BPNS-Trz-FU-C$_{16(H2O)}$.

BP was exfoliated in water (BPNS$_{H2O}$) via tip sonication for 1 h providing a more sustainable and environmentally friendly approach. The resulting material was characterized and subsequently functionalized evaluate its antiviral performance. AFM of BPNS$_{H2O}$ revealed nanosheets with a thickness ranging from 1-10 nm and lateral dimensions of 100-500 nm (Fig. S23I). In the high resolution XPS P2p spectrum, in addition to the characteristic P-P signal at 130.0 eV, a distinct peak at 134.8 eV corresponding to P-O was observed, confirming partial oxidation (Fig.

S23C). Following functionalization with 2-azido-4,6-dichloro-1,3,5-triazine (BPNS-Trz$_{H2O}$) the high resolutionC1s and N1s showed signals characteristic of the triazine moiety (Fig. S23D and E, cf. Fig. 2B-G). The high resolution P2p spectrum further indicated an increased degree of oxidation relative to the BPNS exfoliated in NMP (Fig. S23F and Fig. 2E). ToF-SIMS showed strong PO$_3^-$ fragments for both BPNS$_{H2O}$ and BPNS-Trz$_{H2O}$ indicative of the enhanced oxidation (Fig. S23J and M). Moreover, BPNS-Trz$_{H2O}$ showed an additional CN$^-$ fragment further conforming successful functionalization (Fig. S23O). After functionalization with fucoidan and palmitic acid (BPNS-Trz-FU-C$_{16(H2O)}$), FTIR revealed all characteristic peaks including an OH band at ~3500 cm$^{-1}$, CH vibrations at ~2900 cm$^{-1}$, C=O ester stretching at ~1600 cm$^{-1}$, S=O stretching and deformation at ~1200 cm$^{-1}$ and ~600 cm$^{-1}$ respectively, and C-O ether stretching at 1050 cm$^{-1}$ (Fig. S23P). AFM indicated an increase in sheet thickness (20-100 nm) and lateral dimension (200 nm-2 μm) with a smoother surface morphology (Fig. S23R). Finally, the materials were tested for their biocompatibility. The evaluation of the cytotoxicity demonstrated, that BPNS-Trz-FU-C$_{16(H2O)}$ maintained 80% cell viability at concentrations below 100 μg/mL, confirming low toxicity (Fig. S23S). In plaque-reduction assay, all tested materials (including FU-C$_{16}$, BPNS-Trz-FU$_{(H2O)}$ and BPNS-Trz-FU-C$_{16(H2O)}$) exhibited inhibition relative to the positive control. Notably, BPNS-Trz-FU-C$_{16(H2O)}$ showed the least plaque-formation suggesting, that BP as a platform with aliphatic chains acts as a more effective inhibitor than FU-C$_{16}$ alone. In the virucidal assay, all materials with aliphatic chains achieved complete viral deactivation with no plaque formation. In conclusion, these results demonstrate that the higher oxidation state of BPNS$_{H2O}$ did not impair the antiviral efficacy and that a more environmentally sustainable approach can be applied without decreased activity of the final material.

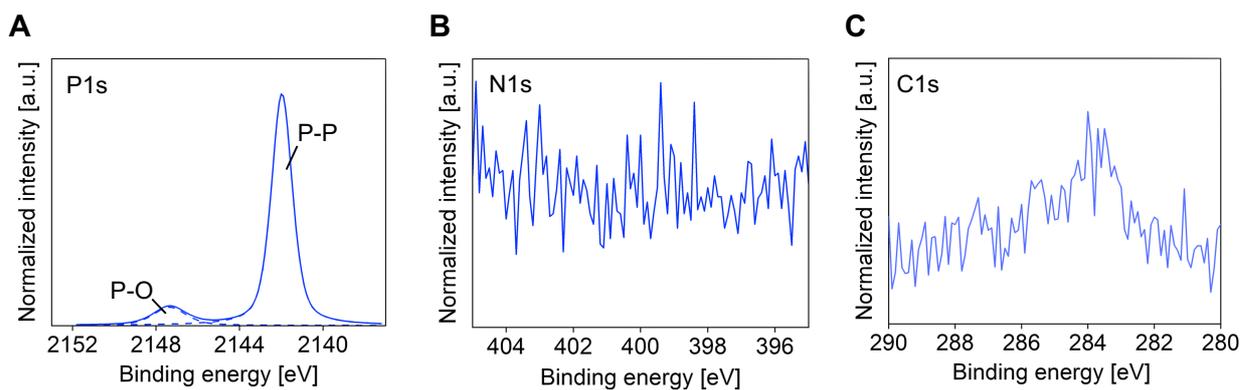

**Figure S4.** High resolution HAXPES spectra with peak fitting for BPNS-Trz-FU-$C_{16}$ (A) P1s, (B) N1s and (C) C1s.

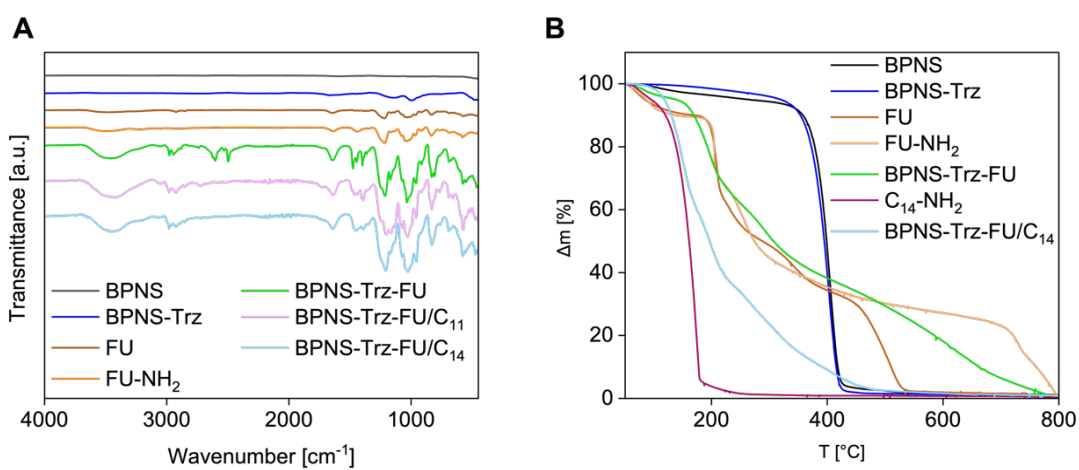

**Figure S5.** Characterization of BPNS-Trz/Cx derivatives including (A) FTIR and (B) TGA.

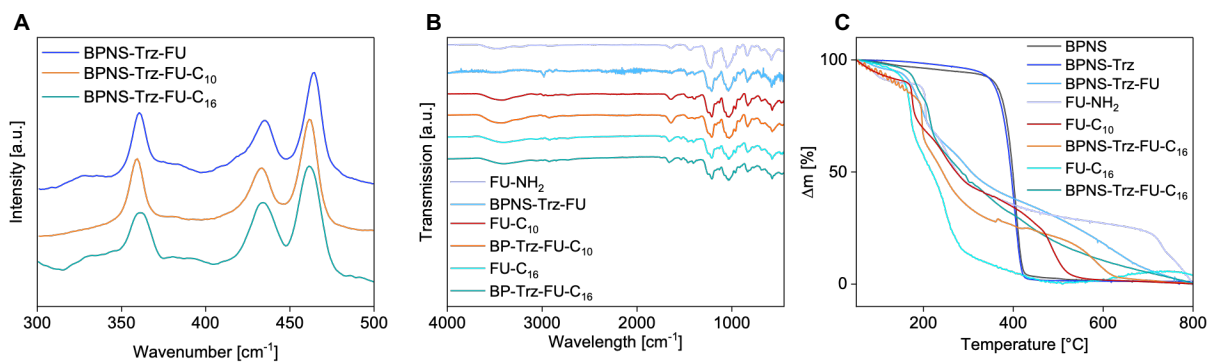

**Figure S6.** Characterization of BPNS-Trz-FU-Cx derivatives including (A) raman, (B) FTIR and (C) TGA.

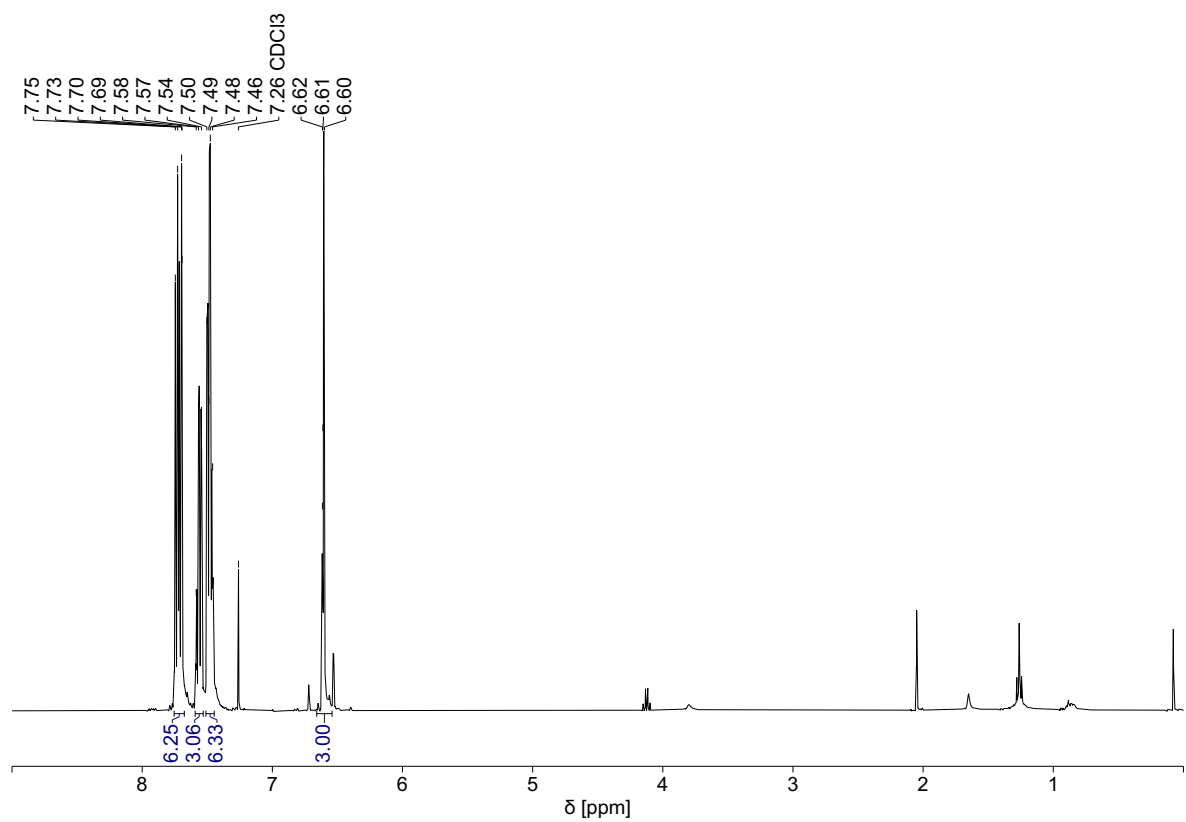

**Figure S7.** $^1$H-NMR of 3,5-dichloro-*N*-(triphenylphosphoranylidene)aniline.

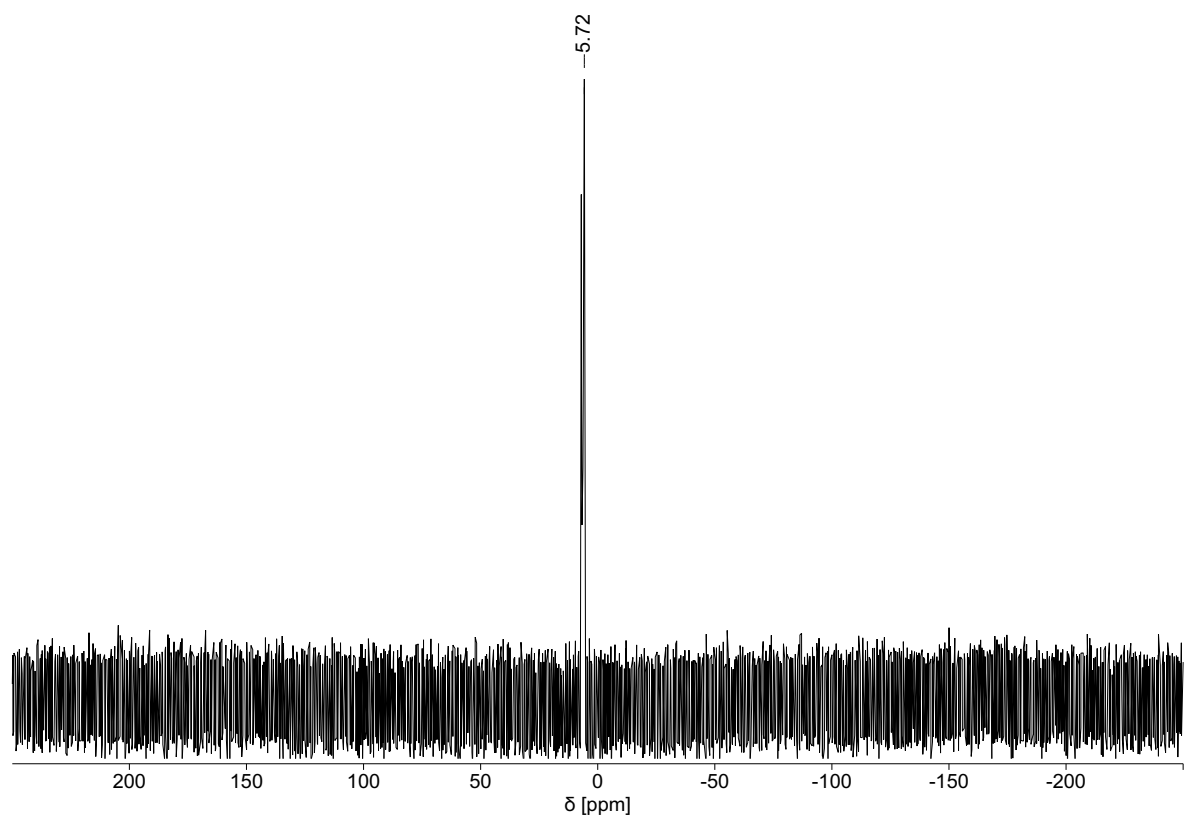

**Figure S8.** $^{31}$P-NMR of 3,5-dichloro-*N*-(triphenylphosphoranylidene)aniline.

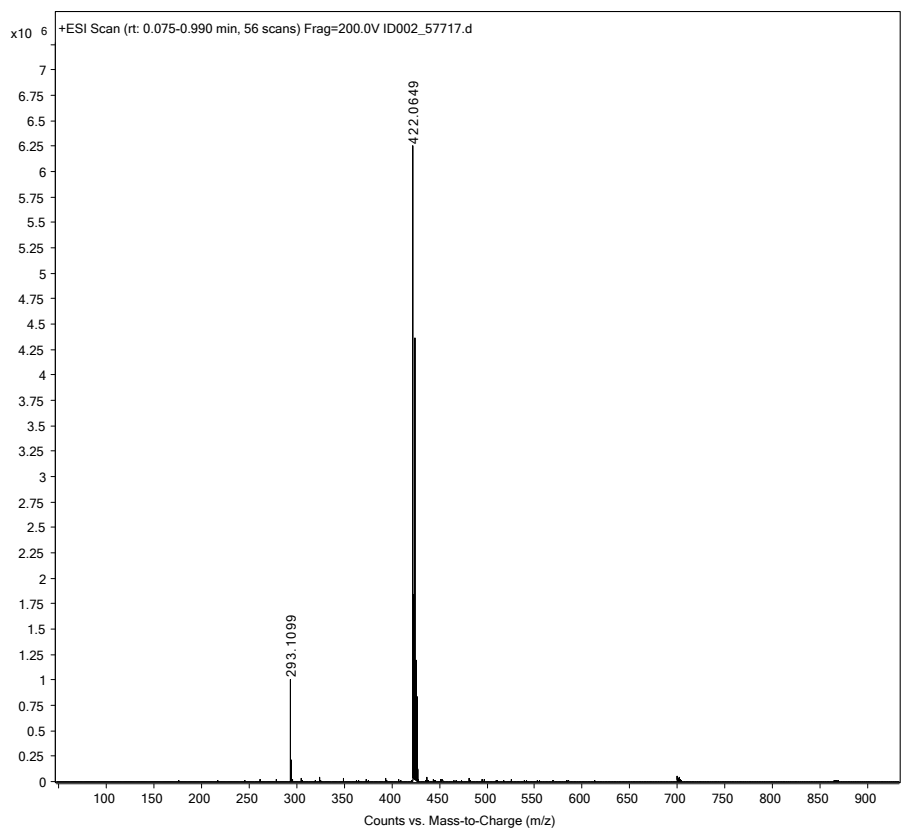

**Figure S9.** ESI of 3,5-dichloro-*N*-(triphenylphosphoranylidene)aniline.

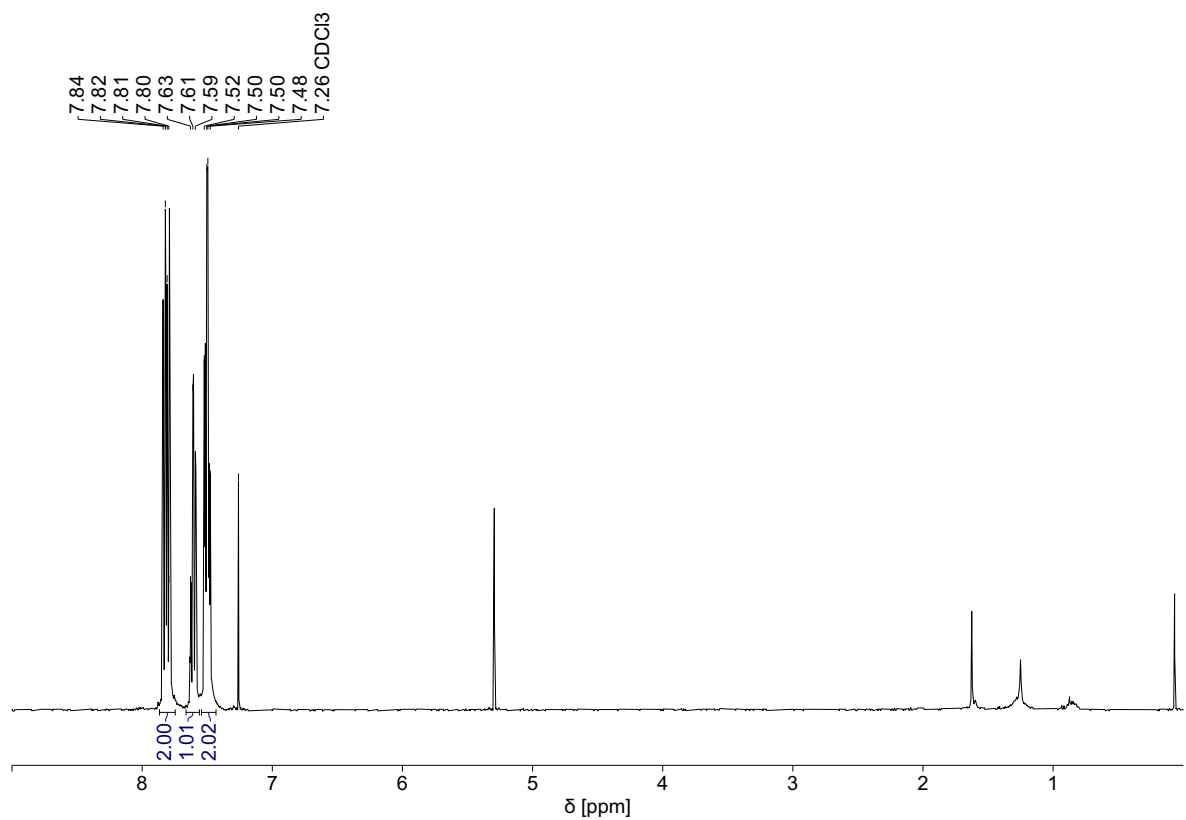

**Figure S10.** ¹H-NMR of *N*-(4,6-dichloro-1,3,5-triazine)triphenylphosphoranylidene.

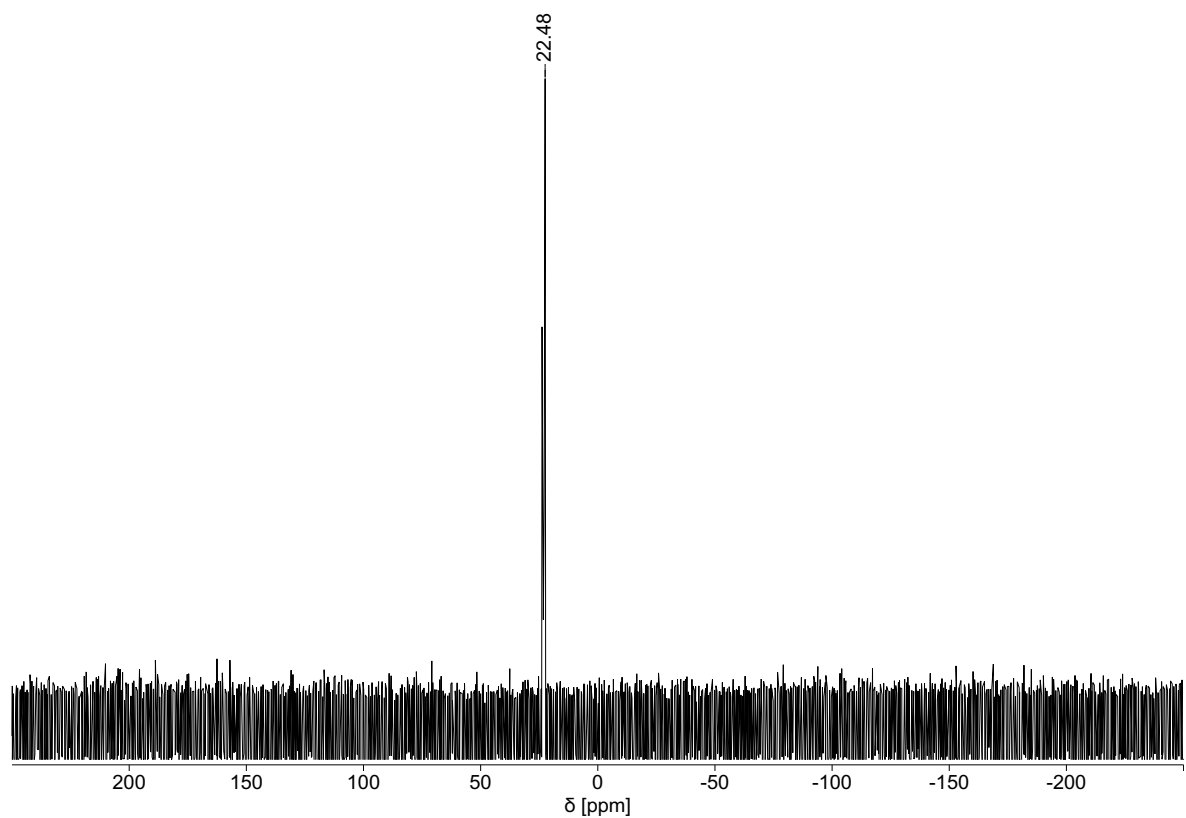

**Figure S11.** $^{31}$P-NMR of *N*-(4,6-dichloro-1,3,5-triazine)triphenylphosphoranylidene.

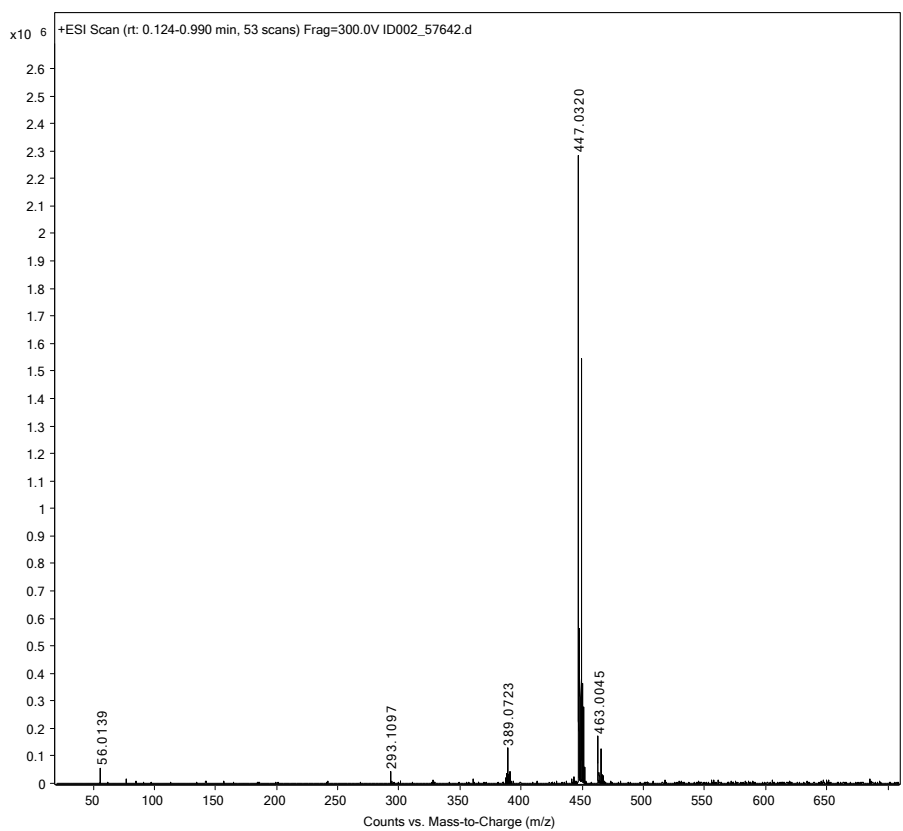

**Figure S12.** ESI of *N*-(4,6-dichloro-1,3,5-triazine)triphenylphosphoranylidene.

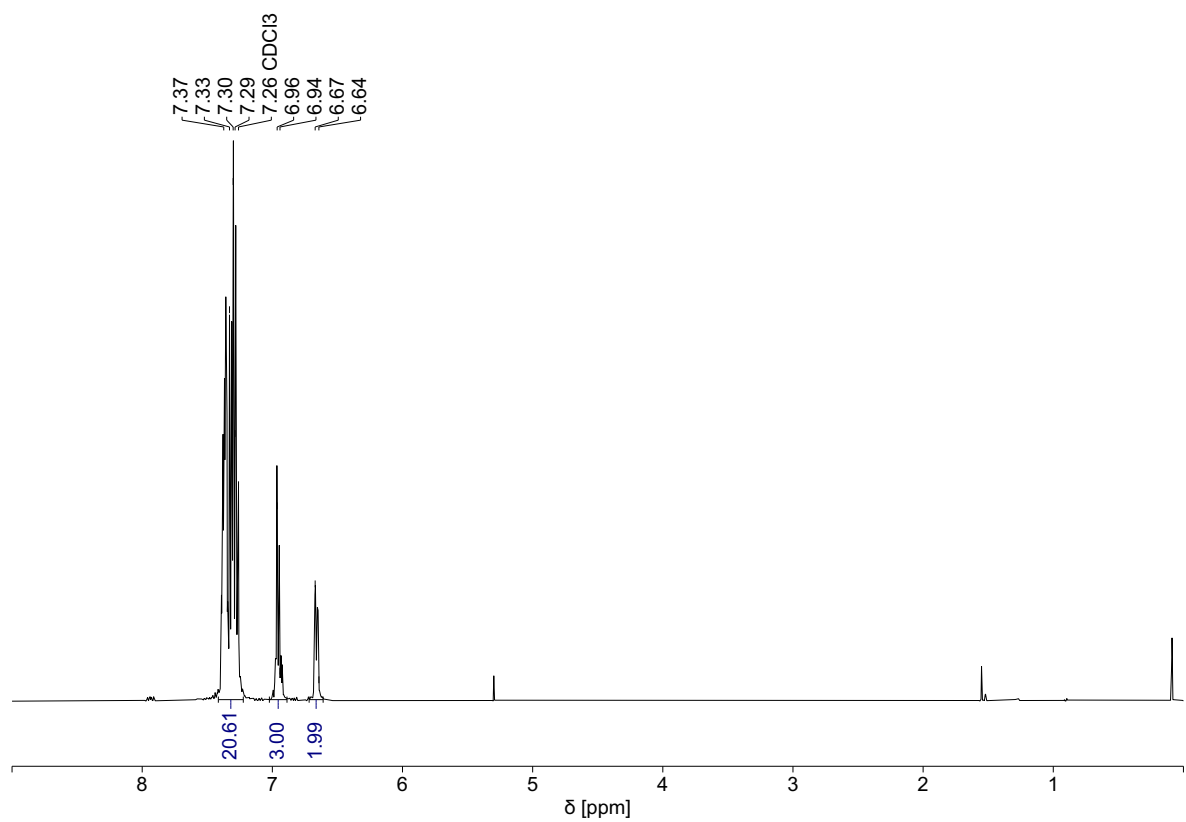

**Figure S13.** ¹H-NMR of N-(diphenylphosphaneyl)-N-1,1-triphenylphosphanamine.

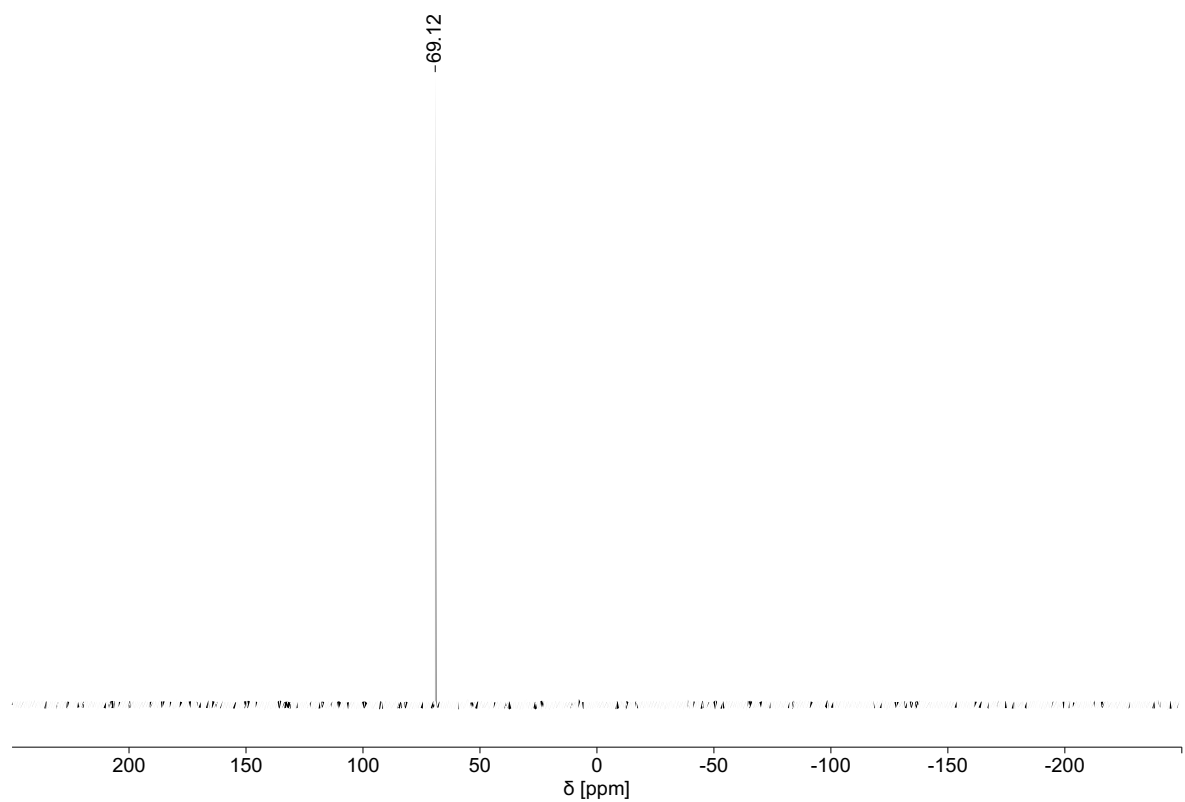

**Figure S14.** ³¹P-NMR of N-(diphenylphosphaneyl)-N-1,1-triphenylphosphanamine.

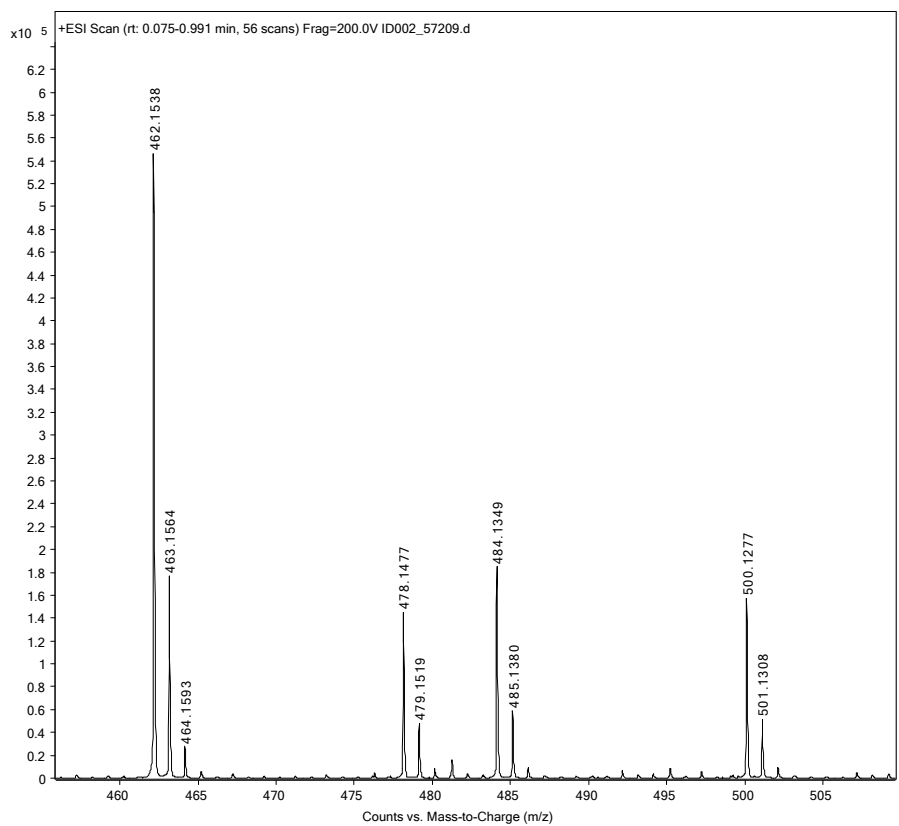

**Figure S15.** ESI of N-(diphenylphosphaneyl)-N-1,1-triphenylphosphanamine.

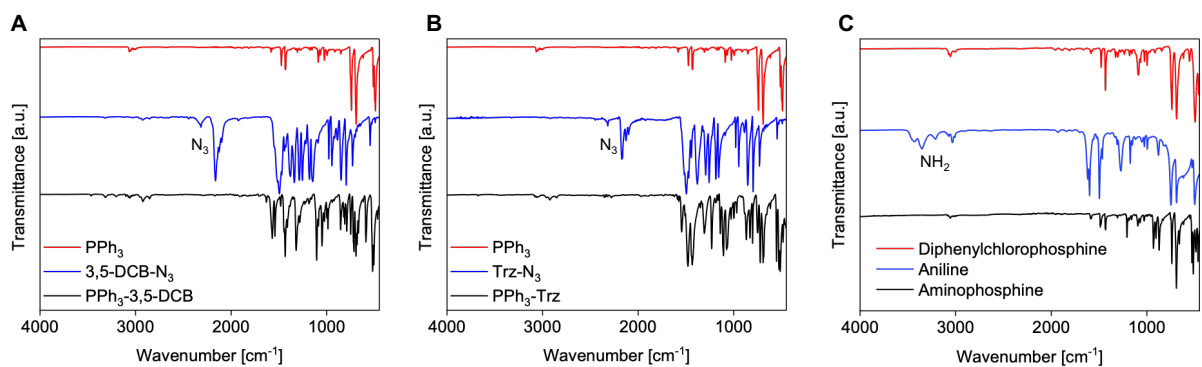

**Figure S16.** IR comparison of synthesized P=N/P-N control materials. Disappearance of the $N_3$ and $NH_2$ signal correspond to successful conversion.